\documentclass[prb,showpacs,twocolumn,amsmath,amssymb,superscriptaddress]{revtex4-1}

\usepackage{graphicx}
\usepackage{dcolumn}
\usepackage{bm}
\usepackage{color}
\usepackage{amsmath}

\newcommand{\cro}{Ca$_2$RuO$_4$}		            

\newcommand{\ttg}{$t_{2g}$}                         
\newcommand{\eg}{$e_{g}$}                           
\newcommand{\dxz}{$d_{xz}$}                         
\newcommand{\dyz}{$d_{yz}$}                         

\begin{document}                  

\title{Magnetic Anisotropy and Orbital Ordering in Ca$_2$RuO$_4$}

\author{D. G. Porter}
\email[]{dan.porter@diamond.ac.uk}
\affiliation{Diamond Light Source Ltd., Harwell Science and Innovation Campus, Didcot, Oxfordshire, OX11 0DE, UK}
\author{V. Granata}
\affiliation{CNR-SPIN, c/o Universit\'a di Salerno- Via Giovanni Paolo II, 132 - 84084 - Fisciano (SA), Italy}
\affiliation{Dipartimento di Fisica ‘E.R. Caianiello’, Universit\'a di Salerno, Fisciano, Salerno I-84084, Italy}
\author{F. Forte}
\affiliation{CNR-SPIN, c/o Universit\'a di Salerno- Via Giovanni Paolo II, 132 - 84084 - Fisciano (SA), Italy}
\affiliation{Dipartimento di Fisica ‘E.R. Caianiello’, Universit\'a di Salerno, Fisciano, Salerno I-84084, Italy}
\author{S. Di Matteo}
\affiliation{Univ Rennes, CNRS, IPR (Institut de Physique de Rennes) - UMR 6251, F-35000 Rennes, France}
\author{M. Cuoco}
\affiliation{CNR-SPIN, c/o Universit\'a di Salerno- Via Giovanni Paolo II, 132 - 84084 - Fisciano (SA), Italy}
\affiliation{Dipartimento di Fisica ‘E.R. Caianiello’, Universit\'a di Salerno, Fisciano, Salerno I-84084, Italy}
\author{R. Fittipaldi}
\affiliation{CNR-SPIN, c/o Universit\'a di Salerno- Via Giovanni Paolo II, 132 - 84084 - Fisciano (SA), Italy}
\affiliation{Dipartimento di Fisica ‘E.R. Caianiello’, Universit\'a di Salerno, Fisciano, Salerno I-84084, Italy}
\author{A. Vecchione}
\affiliation{CNR-SPIN, c/o Universit\'a di Salerno- Via Giovanni Paolo II, 132 - 84084 - Fisciano (SA), Italy}
\affiliation{Dipartimento di Fisica ‘E.R. Caianiello’, Universit\'a di Salerno, Fisciano, Salerno I-84084, Italy}
\author{A. Bombardi}
\affiliation{Diamond Light Source Ltd., Harwell Science and Innovation Campus, Didcot, Oxfordshire, OX11 0DE, UK}
\affiliation{Department of Physics, University of Oxford, Parks Road, Oxford OX1 3PU, United Kingdom}

\date{\today}

\begin{abstract}
We review the magnetic and orbital ordered states in \cro{} by performing Resonant Elastic X-ray Scattering (REXS) at the Ru L$_{2,3}$-edges.
In principle, the point symmetry at Ru sites does not constrain the direction of the magnetic moment below $T_N$. However early measurements reported the ordered moment entirely along the $\vec{b}$ orthorhombic axis.
Taking advantage of the large resonant enhancement of the magnetic scattering close to the Ru L$_2$ and L$_3$ absorption edges, we monitored the azimuthal, thermal and energy dependence of the REXS intensity and find that a canting ($m_c \simeq 0.1 m_b$) along the $\vec{c}$-orthorhombic axis is present. No signal was found for $m_a$ despite this component also being allowed by symmetry. 
Such findings are interpreted by a microscopic model Hamiltonian, and pose new constraints on the parameters describing the model.
Using the same technique we reviewed the accepted orbital ordering picture. We detected no symmetry breaking associated with the signal increase at the ``so-called'' orbital ordering temperature ($\simeq 260$ K). We did not find any changes of the orbital pattern even through the antiferromagnetic transition, suggesting that, if any, only a complex rearrangement of the orbitals, not directly measurable using linearly polarized light, can take place.  

\end{abstract}

\pacs{71.30.+h, 75.50.Ee, 75.30.Gw, 75.25.-j, 75.25.Dk, 75.30.-m, 77.22.Ej, 77.80.B-, 78.70.Ck, 78.70.En}
\maketitle 

\section{Introduction}
 
As it has been known for more than two decades,\cite{Imada1998} magnetic, orbital and lattice degrees of freedom in transition-metal oxides 
can lead to a rich variety of ground states, whose physical interpretation still defies modern research.
Indeed, several phenomena depending on tiny energy differences,\cite{Khomskii2014} often difficult to determine experimentally, can determine a strong coupling of orbital and magnetic degrees of freedom. Among the materials that have been thoroughly investigated in the last twenty years, \cro{} stands as a typical Mott insulator, displaying several phase transitions with temperature. \cro{} is a paramagnetic metal for temperatures above T$_{MI}$=357K,\cite{Alexander1999} below which an insulating, strongly first-order, transition takes place \footnote{Note: Because of an hysteresis cycle, the exact value of the temperature depends on whether it is reached on cooling or on heating}. It then shows, below T$_N$=110K,\cite{Nakatsuji1997} an antiferromagnetic (AFM) transition with magnetic moment along the $\vec{b}$-orthorhombic axis\cite{Braden1998} and $\vec{Q}=0$ propagation vector. Between the two, another phase transition was reported,\cite{Zegkinoglou2005} below T$_{OO} \simeq 260$ K, that was interpreted as due to a ferro-orbital ordering (OO) within the $t_{2g}$ subspace of Ru $4d$ orbitals.
The nature of the insulating state in \cro{}, below T$_{MI}$, has been the subject of intense debate, focused either on the idea of an orbitally selective scenario, where Mott gaps open only on certain orbitals\cite{Anisimov2002,Liu2011,Sutter2017} or on the rejection of such a scenario.\cite{Liebsch2007,Gorelov2010,Zhang2017} In this other view, the orbital selection is not required as the crystal-field splitting, via the compression of the apical RuO bond, is enough to produce a half-filled \dxz{}/\dyz{} insulating state. Yet, the importance of the orbital degree of freedom seems justified by former O K-edge x-ray absorption\cite{Mizokawa2001} and Resonant Elastic X-ray Scattering (REXS) measurements\cite{Zegkinoglou2005} that have observed, respectively, strong variations in the orbital filling with temperature and an apparently second-order phase transition at T$_{OO}$, that cannot be explained through the coupling with the lattice alone.

The aim of the present article is to shed light on the exact trend of the OO with temperature and on the coupling of magnetic and orbital order parameters (OP), through REXS at the Ru L$_{2,3}$-edges. We employ stringent theoretical conditions, described in Section II, that allow a triple projection of both orbital and magnetic degrees of freedom along the three orthogonal orthorhombic axes of the unit cell.
Our main results are the following: 

{\it a)} By exploring the critical behaviour close to the N\'eel ordering we identify unambiguously a non-zero canting of the magnetic moment along the $\vec{c}$-orthorhombic axis. Analogously, we could conclude that the magnetic moment along the $\vec{a}$-orthorhombic axis is practically zero (less than $10^{-2} m_b$).

{\it b)} After disentangling the magnetic OP, it is possible to analyse the behaviour of the orbital OP alone, both around $T_N$ and around T$_{OO}$. At $T_N$, it turns out that there is no significant variation in the square moduli of the orbital filling in the d$_{xy}$ vs. (d$_{xz}$, d$_{yz}$) subspaces. So no population transfer from one subspace to the other. However, the ground state might be affected by changes in the relative phases of d$_{xz}$, d$_{yz}$ and d$_{xy}$ orbitals. This might explain the different behaviour of the (103) reflection at the L$_2$ and L$_3$ edges.
The analysis around T$_{OO}$ confirms the experimental results of Ref. [\onlinecite{Zegkinoglou2005}] at the same (100) reflection. Yet, moving to the off-specular reflection (013), sensitive to the same order parameter (OP), as shown in Section II, no abrupt behaviour at T$_{OO}$ is registered, but rather a continuous increase of the signal from 300K to $T_N$. The latter seems to point rather to a coupling of orbital and lattice degrees of freedom, as explained in Section IV.

The present paper is organized as follows: in Section II the theoretical framework is introduced. Bragg-forbidden reflections are divided in three different classes, each sensitive to a specific component of the magnetic moment. The same classes are also sensitive to the orbital degree of freedom, but with a different azimuthal dependence, allowing a full disentanglement of magnetic and orbital OPs. A complete analysis of all magnetic and orbital components accessible via the corresponding Bragg-forbidden reflections is therefore performed. 
In Section III the experimental setup is described, together with the structural characterization of the sample. Section IV is devoted to the discussion of the results and is divided into three subsections related, respectively, to the disentanglement of orbital and magnetic OP through REXS; to the analysis of the orbital behaviour both around $T_N$ and $T_{OO}$; and, finally, to the description of a theoretical model to explain the antiferromagnetic canting. In Section V we draw our conclusions. We remark that several results can be demonstrated only after some technical analysis. For the sake of a clearer reading, we have postponed most of the technical demonstrations in the four Appendices.

\begin{figure}[ht]
	\centering
	\includegraphics[width=0.4\textwidth]{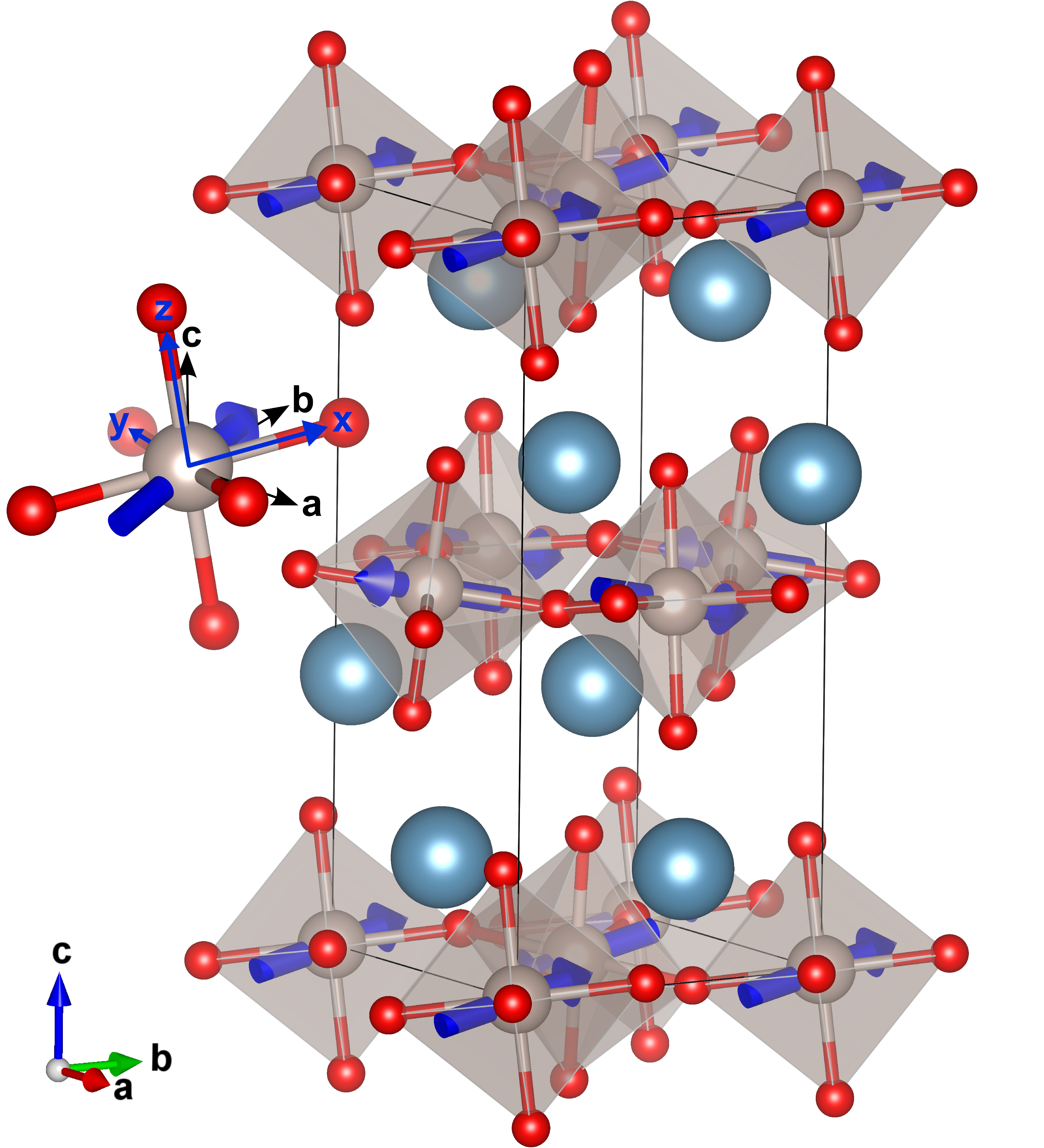}
	\caption{(Colour online) Crystal structure of \cro{}. Grey spheres indicate the canted octahedral Ru sites, light blue spheres are calcium and red are oxygen. Blue arrows indicate the magnetic moment on each Ru site. The local $xyz$ frame around Ru$_1$ is highlighted in the inset.}\label{fig1}
\end{figure}

\section{Theoretical framework}

The space group of \cro{} is {\it Pbca}, with 4 Ru atoms per unit cells at {\it 4a} positions.\cite{Braden1998} In all the insulating phases, below T$_{MI}$, the symmetry at the three atomic sites Ru$_2$=(0.5,0,0.5), Ru$_3$=(0,0.5,0.5) and Ru$_4$=(0.5,0.5,0) is related to the one of Ru$_1$=(0,0,0) by the two-fold rotations ${\hat{C}}_{2}$ (around the c-axis), ${\hat{B}}_{2}$ (around the b-axis) and ${\hat{A}}_{2}$ (around the a-axis), respectively. In the AFM phase, the {\it Pbca} {\it magnetic} space group (N. 61.1.497 of [\onlinecite{Litvin2013}]) puts the following constraints on the magnetic components of the four atoms: $\vec{m}_1=(m_a,m_b,m_c)$, $\vec{m}_2=(-m_a,-m_b,m_c)$, $\vec{m}_3=(-m_a,m_b,-m_c)$ and $\vec{m}_4=(m_a,-m_b,-m_c)$.
Neutron measurements \cite{Braden1998} point to $m_a=m_c=0$. However, an intriguing remark is that such a condition does not follow from any symmetry requirement. The symmetry constraints of the {\it Pbca} magnetic space group allow for both $m_a$ and $m_c$ different from zero, provided the four magnetic moments in the unit cell are related to one another by the above relations. In this respect, the experimental findings $m_a=m_c=0$ are puzzling: is there a hidden symmetry that forbids the moment from pointing in $a$ and $c$ directions? Or, rather, is the $\vec{m}=(0,m_b,0)$ direction determined by dynamical electronic interactions? If this were the case, we would rather expect a canting of the magnetic moment out of the $m_b$ direction. We address this question in Sections IV.A and IV.C.
 
We also notice that another magnetic structure was reported \cite{Braden1998} in \cro{}. This alternative structure, not observed in our sample, is characterized by the same propagation vector, but belongs to a different magnetic space group ({\it Pbc'a'}, N. 61.4.500 of [\onlinecite{Litvin2013}] with a different axis choice) and was reported to occur at different temperature. The alternative structure is stabilized in presence of a different stoichiometry or slight doping with La, or Ti,\cite{Kunkemoller2015,Kunkemoller2017,Pincini2018} but it should be considered as a different case, as it belongs to a different magnetic irreducible representation of the high-temperature parent symmetry, {\it Pbca1'}. In fact, recently a slight ferromagnetic canting of the moment toward the a-axis was reported to occur in this case,\cite{Kunkemoller2017} that would be forbidden in {\it Pbca}. The reason why we can definitely exclude the magnetic symmetry {\it Pbc'a'} in our case is discussed at the end of Appendix A.

The presence or not of the $m_a$ and $m_c$ components, as well as the precise determination of the OO throughout the insulating phase, can be definitively settled by a complete REXS experiment, looking at all components of the magnetic moment and all accessible components of the OO. The theoretical framework is the following: in resonant conditions, the atomic scattering factor $f_i$ ($i=1$ to $4$) associated to each Ru atom becomes a tensor, $f_i^{\alpha\beta}$ (with $\alpha$ and $\beta$ Cartesian components along the three crystallographic axes $a$, $b$, $c$), as detailed in Appendix A. For this reason the structure factor at the Ru L-edges, in both the PM and the AFM insulating phases, can be written as:

\begin{align}
F_{hkl} &= f_1 + (-)^{h+l} f_2 + (-)^{k+l} f_3 + (-)^{h+k} f_4 \nonumber \\ 
&= \left(1 + (-)^{h+l}{\hat{C}}_{2}\right) \left(1 + (-)^{k+l}{\hat{B}}_{2} \right) f_1^{\alpha\beta}
\label{eq1}
\end{align} 

From Eq. (\ref{eq1}) we deduce that there are three classes of Bragg-forbidden reflections that become allowed at L$_{2,3}$-resonances, each sensitive to a different component of the magnetic moment and to a different projection of the OO (measured by the electric quadrupole components $Q_{\alpha\beta}$, defined in Appendix B). In order to find out the $t_{2g}$ and $e_g$ orbital components associated to each $Q$ tensor, we need to rotate these tensors to the local frame centred at the Ru$_1$ site and directed towards the surrounding oxygen octahedron (frame xyz in Fig. \ref{fig1}).
With the conventions developed in Appendices A and B, we find the values reported in Table \ref{tab:ref_sym}.

\begin{table}[ht]
\centering
\begin{tabular}{c|ccc}
       & $k+l=$even  & $k+l=$odd   & $k+l=$odd  \\
       & $h+l=$odd  & $h+l=$even  &  $h+l=$odd  \\
			 &  e.g. (013), (100)        &  e.g. (103), (010)        &  e.g. (110), (003)       \\ \hline 
$m_a$  & 0 & 0 & 1 \\
$m_b$  & 1 & 0 & 0 \\
$m_c$  & 0 & 1 & 0 \\ \hline
$Q_{ab}$  & 0 & 1 & 0 \\
$Q_{ac}$  & 1 & 0 & 0 \\
$Q_{bc}$  & 0 & 0 & 1 \\ \hline
$|d_{xy}|^2$  & 0.04 & 0.16 & 0.00 \\
$|d_{xz}|^2$  & 0.24 & 0.03 & 0.66 \\
$|d_{yz}|^2$  & 0.60 & 0.00 & 0.29 \\
$|d_{3z^2-r^2}|^2$ & 0.12 & 0.00 & 0.01 \\
$|d_{x^2-y^2}|^2$  & 0.00 & 0.81 & 0.04 \\
\end{tabular}
\caption{Sensitivity of different reflections to magnetic ($m_{\alpha}$) and orbital ($Q_{\alpha\beta}$) directions, from Eq. (\ref{eq1}). The $d$-orbital sensitivity is deduced from $Q_{\alpha\beta}$ through Eq. (\ref{eq2bis}).}
\label{tab:ref_sym}
\end{table}

We remind that magnetic dipoles necessarily rotate the polarization of scattered X-rays, whereas OO also radiates in the nonrotated $\sigma\sigma$ channel, except at on-axis reflections related to the two-fold rotation axis, like the (100) or the (003), where the $\sigma\sigma$ channel is identically zero, as calculated in Appendix C. For this reason, it is possible to select angles and/or polarization channels at which specific orbital or magnetic components are allowed, in this way investigating just that component. The full list of azimuthal scans is given in Appendix C.

\section{Experimental setup and sample structural characterization.}

Single crystals of \cro{} were grown using a floating zone furnace and characterised using a molybdenum source supernova diffractometer (Oxford Diffraction) with a cryojet (Oxford Instruments) for temperature control. Plate shaped samples were cleaved from the grown boule, typically having dimensions $\approx 1000\times1000\times100 ~\mu m^3$ and the (00L) direction perpendicular to the large face.  Several samples were measured on the materials and magnetism beamline, I16 at Diamond Light Source Ltd. Measurements were performed at the Ru $L_2$ and $L_3$ absorption edges (2.967 keV and 2.828 keV, respectively) by reconfiguring the beamline for low energies; performing four bounces on the silicon monochromator, minimising the air path and extending the area detector capability to low energies. To maximise the possible azimuthal rotation range for reflections of interest, samples were mounted in the vertical geometry, with either the (001), (100) or (110) directions perpendicular to the natural polarisation plane of the incident x-ray beam, such that the scattering conditions for reflections along (00L), (H00) or (HH0), respectively, were close to the rotation axis of the diffractometer. In all cases the azimuthal zero angle was defined along the (010) reflection. When mounted in the (100) or (110) directions, this meant scattering from the edge of a sample, with a sample surface \textless $100\times100 ~\mu m^2$. In either case, it was not possible to polish the surface without damaging the sample, as such the surface was scanned for optimal diffraction intensity. The incident energy was set at 2.828 keV or 2.967 keV and was scanned 40 eV around each absorption edge in 0.5 eV steps matching the resolution of the instrument. At this energy, the focused spot size was $\approx180\times50 \mu m^2$. The polarisation of the diffracted beam was analysed by rotating the scattering plane of a highly oriented (002) graphite plate. The cross-channel leakage of the analyser crystal at this energy was \textless 5\%.

A single crystal sample was measured continuously between 90K and 400K using a temperature controlled X-ray diffractometer, at each temperature, full coverage to a high angle of reciprocal space was obtained and refinements were performed to determine the octahedral bond lengths at each temperature. The refined structure parameters for 90K, 150K, 300K and 400K  are given in Table \ref{tab:xrd} and the temperature dependence is shown in Fig. \ref{fig:xrd}. Crystal structures for these and additional temperatures are included in the supplementary material\footnote{See See Supplemental Material at [URL will be inserted by publisher] for [CIF files from x-ray structure refinements].}. These characterisation results are in agreement with the literature on this compound.\cite{Braden1998}

\begin{table}[htp]
    \centering
       \begin{tabular}{c|cccc} \hline
    & 90K & 150K & 300K & 400K \\ \hline
a & 5.3831(3) & 5.3906(3) & 5.4047(3) & 5.3569(3) \\
b & 5.6318(4) & 5.6247(4) & 5.5089(4) & 5.3469(4) \\
c & 11.7288(6) & 11.7332(6) & 11.9130(6) & 12.2575(6) \\
Volume & 355.58(4) & 355.76(4) & 354.70(4) & 351.09(4) \\
 x$_{Ca}$ & 0.00309(6) & 0.00338(7) & 0.00779(6) & 0.00983(9) \\
 y$_{Ca}$ & 0.05929(7) & 0.05861(8) & 0.04665(8) & 0.02565(8) \\
 z$_{Ca}$ & 0.35245(3) & 0.35243(3) & 0.35106(3) & 0.34861(4) \\
 x$_{O1}$ & 0.1940(2) & 0.1945(2) & 0.1969(2) & 0.1928(3) \\
 y$_{O1}$ & 0.3016(2) & 0.3011(2) & 0.3013(2) & 0.3071(3) \\
 z$_{O1}$ & 0.0277(1) & 0.0278(1) & 0.0245(1) & 0.0151(1) \\
 x$_{O2}$ & -0.0696(3) & -0.0699(3) & -0.0605(3) & -0.0388(3) \\
 y$_{O2}$ & -0.0220(2) & -0.0213(3) & -0.0178(2) & -0.0092(3) \\
 z$_{O2}$ & 0.1645(1) & 0.1645(1) & 0.1647(1) & 0.1655(1) \\
RuO1 & 2.0202(13) & 2.018(1) & 1.993(1) & 1.951(2) \\
RuO2 & 1.969(1) & 1.970(2) & 1.992(2) & 2.040(2) \\
O1RuO1 & 88.99$^{\circ}$(5) & 89.13$^{\circ}$(5) & 90.16$^{\circ}$(5) & 90.60$^{\circ}$(7) \\
O1RuO2 & 89.62$^{\circ}$(6) & 89.56$^{\circ}$(6) & 89.12$^{\circ}$(6) & 88.85$^{\circ}$(6) \\
RuO1Ru & 149.56$^{\circ}$(1) & 149.69$^{\circ}$(7) & 151.09$^{\circ}$(7) & 152.1$^{\circ}$(1) \\
$\phi$ & 15.2196$^{\circ}$(4) & 15.16$^{\circ}$(4) & 14.46$^{\circ}$(4) & 13.96$^{\circ}$(5) \\ 
 R$_w$(all) & 2.43\% & 2.50\% & 2.47\% & 2.45\% \\
        \end{tabular}
			\caption{Refined structure parameters from single crystal XRD measurements using an Mo-source diffractometer. RuO1 and RuO2 distances are in \AA. O1 and O2 labels refer to Fig. \ref{fig:distortion} and $\phi$ is the out-of-plane rotation of the RuO$_6$ octahedra. CIF files with full refinement details are available in the supplementary material.}
    \label{tab:xrd}
\end{table}

\begin{figure}[ht]
	\centering
	\includegraphics[width=0.5\textwidth]{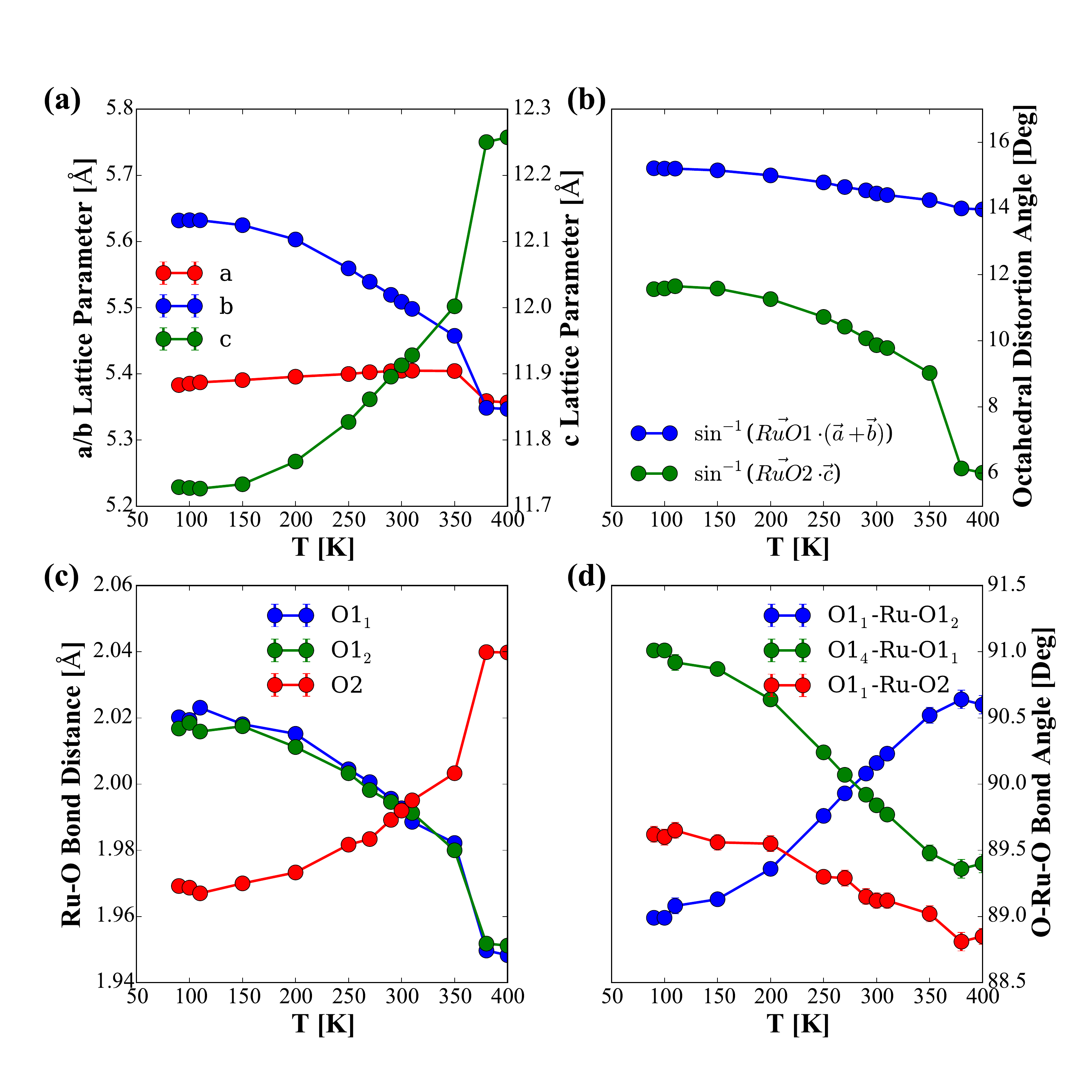}
	\caption{(Colour online) Temperature variation of refined structure parameters from single crystal XRD measurements. (a) Lattice parameters, with the c-axis on a separate scale. (b) Octahedral distortions angles, as defined in Fig. \ref{fig:distortion}. Panels (c) and (d) show the change in Ru-O bond distance and bond angles, respectively. Oxygen ions are labelled as in Fig. \ref{fig:distortion}.}\label{fig:xrd}
\end{figure}

\begin{figure}[ht]
	\centering
	\includegraphics[width=0.4\textwidth]{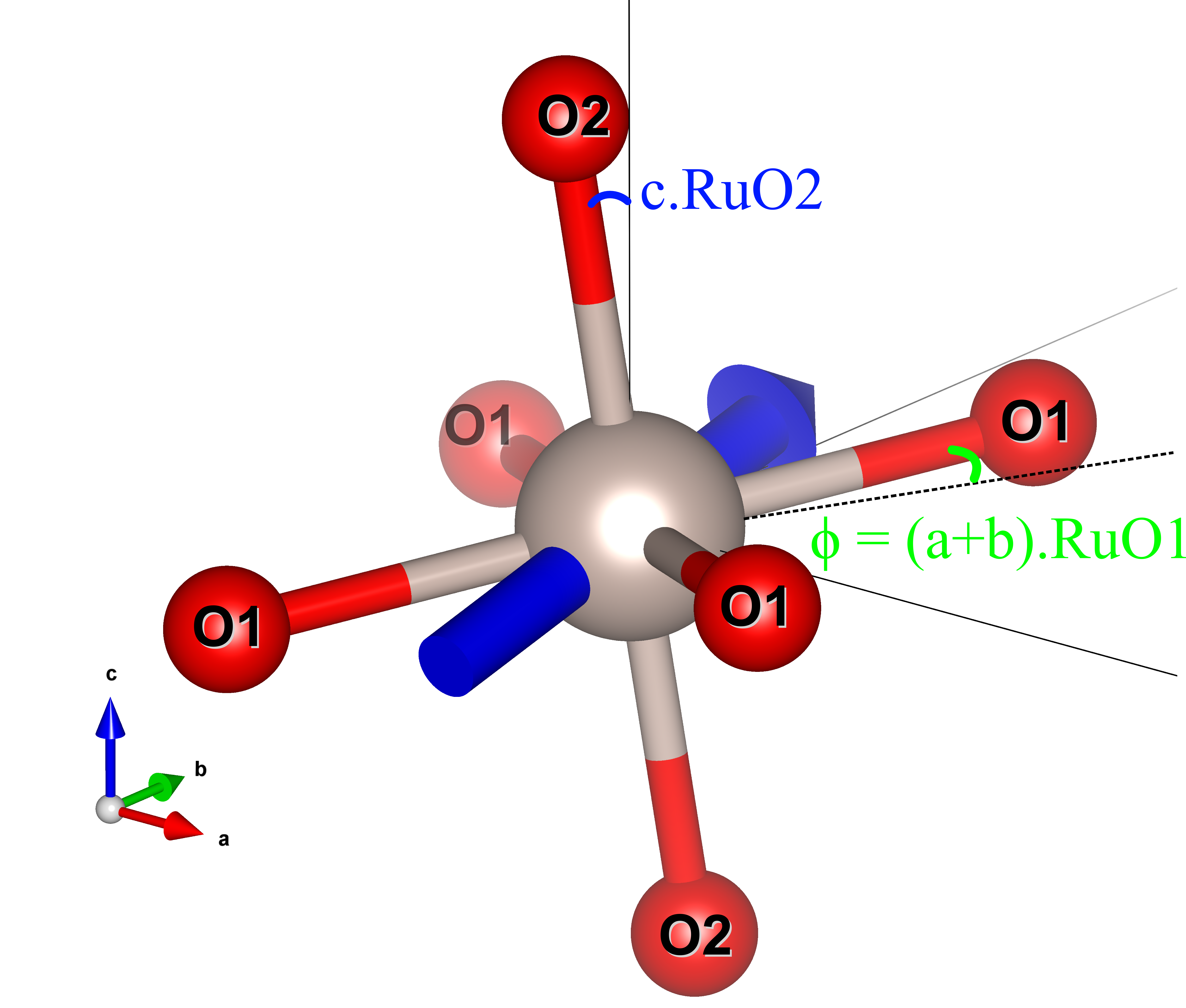}
	\caption{(Colour online) RuO$_6$ octahedra labelling the oxygen atoms and bond angles for Fig. \ref{fig:xrd} and for Table \ref{tab:xrd}.}\label{fig:distortion}
\end{figure}

As previously noted in Ref. [\onlinecite{Braden1998}], at low temperature the Ru-O octahedra are distorted away from the principle axes of the system, with the Ru-O2 bonds, pointing roughly along the $c$-axis (see Fig. \ref{fig:distortion}), shorter than the roughly in-plane Ru-O1 bonds. However as temperature increases the difference between the bond lengths decreases until they become equal just below 300K (see Fig. \ref{fig:xrd}(c)). Above this temperature the Ru-O2 bonds lengthen and become much longer than the Ru-O1 bonds in the high temperature phase. It is worth noting that there is no significant change in the Ru-O1 bond distance across the MI transition and that this mostly occurs in the apical Ru-O2 bonds.

The RuO octahedron keeps its symmetric shape throughout the temperature range. The octahedron rotates with temperature, both away from the c-axis and within the plane as illustrated in Fig. \ref{fig:xrd}(b) by the octahedral distortion angle - defined as the angle between apical Ru-O2 and the $c$-axis, see Fig. \ref{fig:distortion}. At low temperature this angle is greater than 10 degrees but decreases with increasing temperature before a large reduction across the MI transition.

The refined measured structure factors here are in good agreement with those in previous reports.\cite{Braden1998,Friedt2001,Steffens2005} Significant variations are therefore happening to the atomic structure across this temperature range and this must be taken into account to address the changes observed in this system.

\section{Results and discussion}

\subsection{Disentangling orbital and magnetic signals}

Following the theoretical scheme developed in Section II, together with the calculations of Appendices A, B, C, we analyzed the three classes of reflections reported in Table \ref{tab:ref_sym}.
We first remark that it is not possible to disentangle the orbital and magnetic OPs, $m$ and $Q$, at on-axis reflections. For example, because of the geometry, at the (100) reflection, we would have (Eq. \ref{eq2c}) $I_{\sigma\sigma}=0$ identically, and $I_{\sigma\pi}= (m_b^2 + Q_{ac}^2)\cos^2\theta_B \cos^2\psi $, with the same azimuth dependence for the magnetic and orbital OPs. In particular, this does not allow to attribute the intensity increase below $T_N$ to a magnetic contribution or rather to the onset of an enhanced OO in correspondence to the magnetic transition.\cite{Wilkins2005} In order to double check this point, we need to move to off-axis reflections. 

\begin{figure}[ht]
	\centering
	\includegraphics[width=0.4\textwidth]{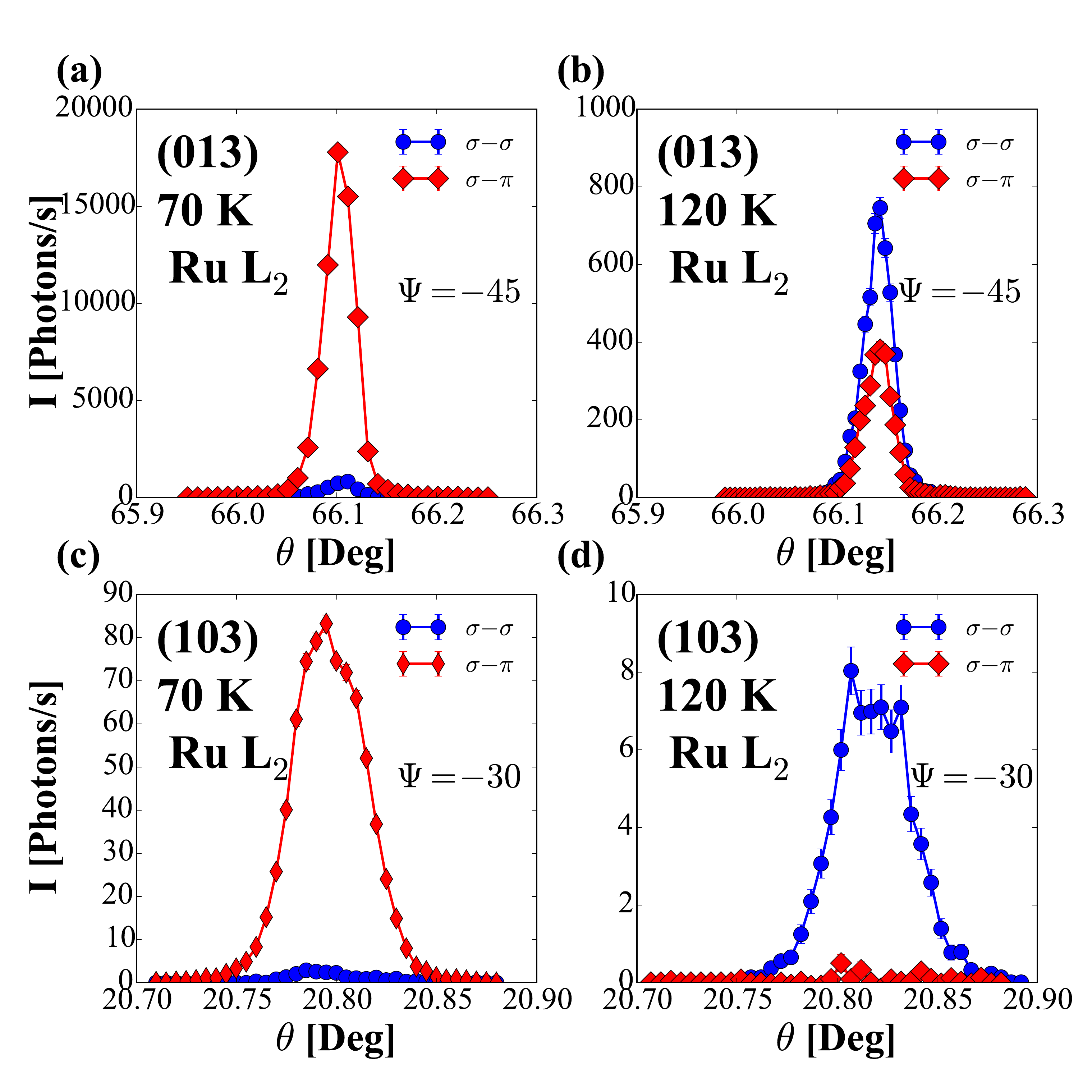}
	\caption{(Colour online) Rocking curve of reflections (013) and (103) at 2.967 keV, above and below T$_N$. Crosstalk between the polarisation channels has been removed in each case and intensities are corrected for self-absorption.}
	\label{fig3}
\end{figure}

\begin{figure}[ht]
	\centering
	\includegraphics[width=0.4\textwidth]{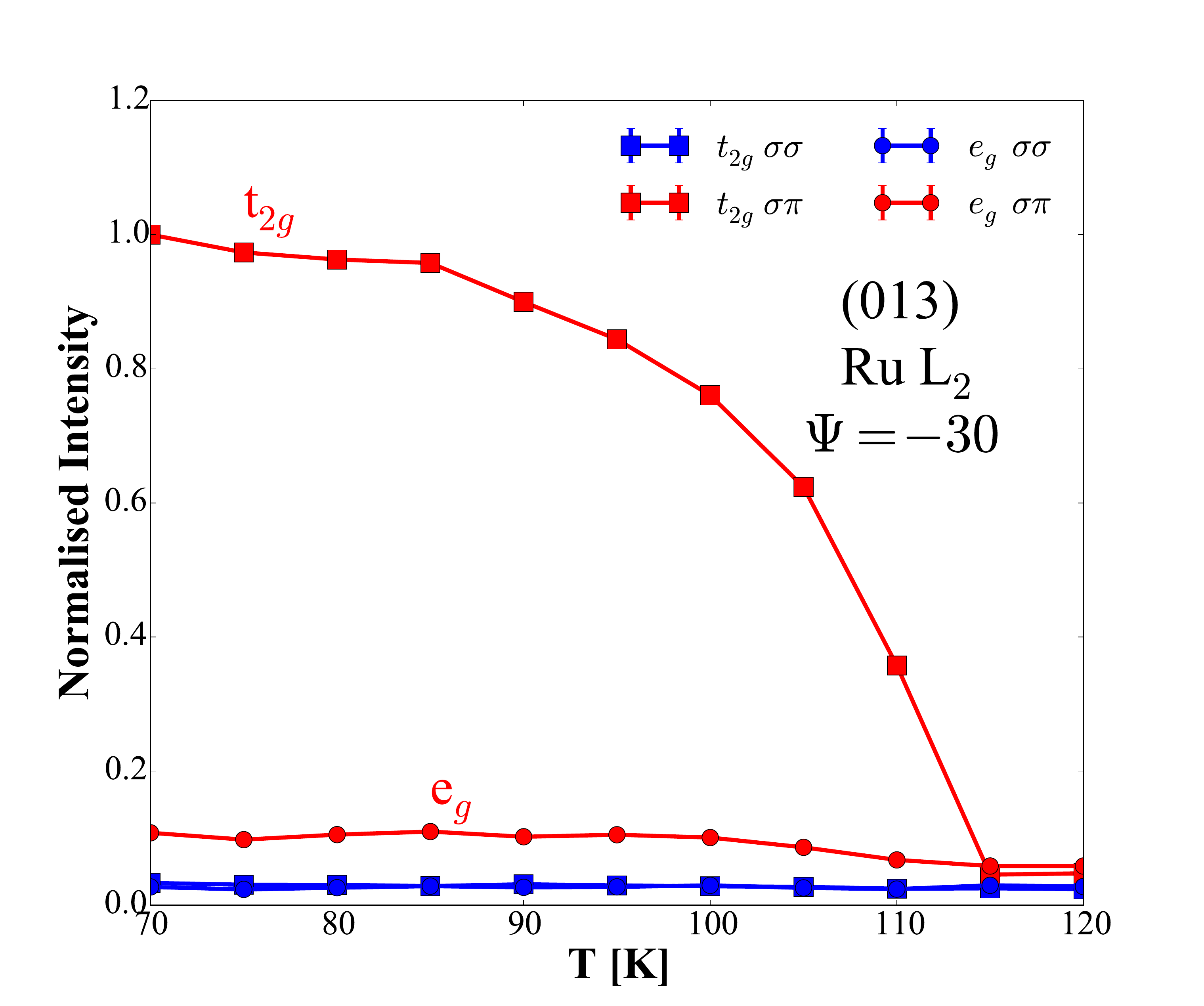} 
	\caption{(Colour online) Temperature dependence of the $(013)$ reflection, showing the absence of any practical change at $T_N$ in the $\sigma\sigma$ channel. This implies that the corresponding orbital OP is unchanged at $T_N$.} \label{fignoOO}
\end{figure}

{\it Case $m_{b}$ and $Q_{ac}$ -} In Figure \ref{fig3}(a) and (b) we represent the rocking curve of the (013) reflection at the L$_2$ edge, above and below $T_N$, in both $\sigma\sigma$ and $\sigma\pi$ channels at the azimuth $\psi = -45^{\circ}$ (we defined the azimuth $\psi = 0^{\circ}$ when the scattering plane contains the b-axis with positive projection of the outgoing wave-vector on it). In this case, we get the following theoretical expressions (Eq. \ref{eq1c}): 
$I_{\sigma\pi}=0.65 m_b^2 + 0.19 Q_{ac}^2$ and $I_{\sigma\sigma}=0.32 Q_{ac}^2$. 
Above $T_N$, the theoretical ratio $I_{\sigma\sigma}/I_{\sigma\pi} \simeq 1.7$ compares well with the experimental ratio $\sim 1.75$ of Fig. \ref{fig3}(b). Below $T_N$, we find an increase in $I_{\sigma\pi}$ by a factor $\sim 40$, in correspondence of no practical increase in $I_{\sigma\sigma}$. As scattering in $\sigma\sigma$-channel can only arise from quadrupolar terms,\cite{sdm2012} this means that the whole variation in intensity below T$_N$ is magnetic.
The absence of any variation in the orbital signal can be also extracted from the practically flat behaviour of the $\sigma\sigma$ intensity in temperature through the Neel transition and down to the lowest temperatures, as shown in Fig. \ref{fignoOO}.

{\it Case $m_{c}$ and $Q_{ab}$ -} In Figure \ref{fig3}(c) and (d) we represent the rocking curve of the (103) reflection at the L$_2$ edge, above and below $T_N$, in both $\sigma\sigma$ and $\sigma\pi$ channels at the azimuth $\psi = -30^{\circ}$. Using Eq. \ref{eq3c}, we get the following theoretical expressions (within a common multiplicative constant): $I_{\sigma\pi}=0.57m_c^2 + 2\cdot 10^{-3} Q_{ab}^2$ and $I_{\sigma\sigma}=0.49 Q_{ab}^2$. Above $T_N$, the experimental $\sigma\pi$ signal is practically zero, confirming the above expression, as $m_c$ is zero above $T_N$. The $\sigma\sigma$ intensity shows that $Q_{ab}$ is not zero and also that it does not change appreciably throughout the magnetic transition. This is interesting because, when combined to the same behaviour of the $\sigma\sigma$ (013) reflection (see Fig. \ref{fignoOO}) and with Table I, it shows that no orbital rearrangement takes place at $T_N$ in the square modulus of each $t_{2g}$ orbital occupations. We can therefore attribute unambiguously the increase by a factor of more than 600 in the $\sigma\pi$ channel to the onset of a non-zero $c$ component of the magnetic moment. The relative ratio of the intensity at the (013), for $m_b$ and at the (103), for $m_c$, provides the following ratio for the two components: $m_c \sim 0.1 m_b$.

\begin{figure}[ht]
	\centering
	\includegraphics[width=0.4\textwidth]{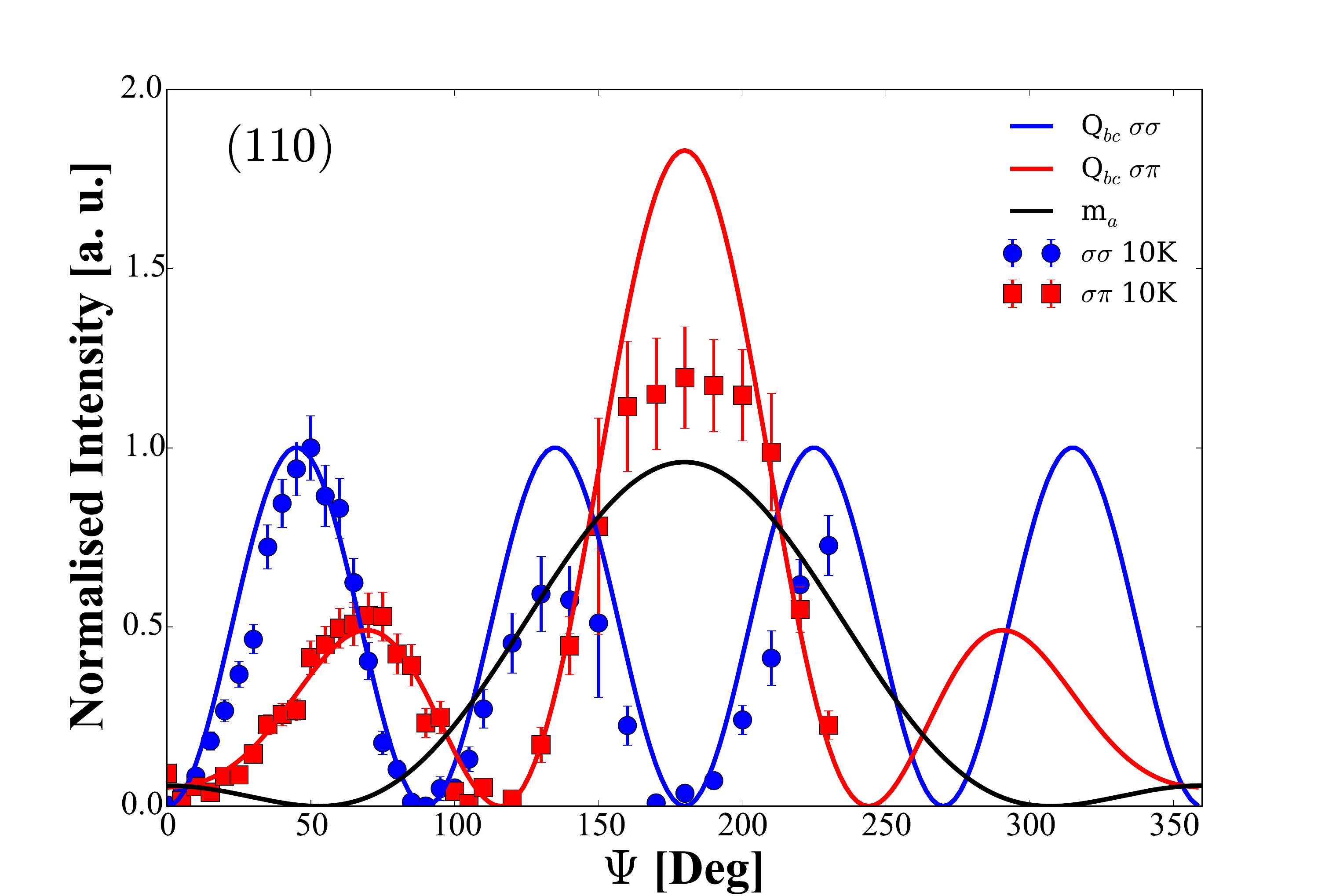} 
	\caption{(Colour online) Azimuthal scan of the $(110)$ reflection, at 10K, that can be described only by the quadrupole term $Q_{bc}$, in both $\sigma\sigma$ and $\sigma\pi$ channels, as shown by the ab-initio FDMNES calculations represented here as blue and red continuous lines. This points to the absence of any magnetic component $m_a$, whose theoretical contribution from Eq. \ref{eq5c} is also reproduced here, as a black continuous line.} 
	\label{fignoma}
\end{figure}

{\it Case $m_{a}$ and $Q_{bc}$ -} The third class of reflections, sensitive to $m_a$ and $Q_{bc}$, is analyzed at the (110) reflection, as shown in Fig. \ref{fignoma}. In the $\sigma\pi$ channel the magnetic and the quadrupole terms have different azimuthal scans. Based on symmetry considerations alone (see Eq. (\ref{eq5c})), we can write their amplitudes as follows:
$0.37 \cos(2\psi)+0.61 \cos\psi$ for Q$_{bc}$, and $0.37 +0.61 \cos\psi$ for m$_a$. As shown in Fig. \ref{fignoma}, the zero of the experimental data in the $\sigma\pi$ data around $120^\circ$ cannot be explained if the magnetic contribution is present. So, only $Q_{bc}$ is present in the spectrum, both above and below $T_N$. This is further confirmed, independently, by a fit with both $m_a$ and $Q_{bc}$ terms, that gives $0\%$-weight for $m_a$, and also by an independent ab-initio simulation through the FDMNES program,\cite{Bunau2009} that allowed to describe the experimental data fairly well with no magnetic components, as shown in Fig. \ref{fignoma}. We can conclude therefore that the component $m_a=0$.

\subsection{Orbital behaviour around $T_N$ and $T_{OO}$.}

The above discussion highlighted the absence of detectable changes around T$_N$ for the orbital degrees of freedom. Yet, we should remind, from Table I, that in our experiments we are sensitive only to the square moduli of the orbital fillings. This means that we can only state that no spectral-weight transfer takes place, e.g., from $d_{xy}$ subspace to ($d_{xz}$, $d_{yz}$) subspace. However, we cannot exclude that a readjustment of phases takes place with the onset of complex orbital ordering (e.g., $d_{xz}\pm i d_{yz}$), driven by the spin-orbit coupling. Such a readjustment might for example explain the increase in the low-temperature peak associated to the apical oxygen ions in the O K-edge x-ray absorption experiment of Ref. [\onlinecite{Mizokawa2001}], that was performed with circular polarization. In fact, only circular polarization is sensitive to complex linear combinations. Specific phase relations within the $t_{2g}$ manifold might also describe the relative changes between the L$_2$ and L$_3$ channels (see Appendix D), analogously to what happens in iridates, where, {\it mutatis mutandis}, L$_2$ and L$_3$ signals are different below $T_N$, because a specific, $j=5/2$, linear combination sets in within the $t_{2g}$ manifold.\cite{kim2009} 

\begin{figure}[ht]
	\centering
	\includegraphics[width=0.4\textwidth]{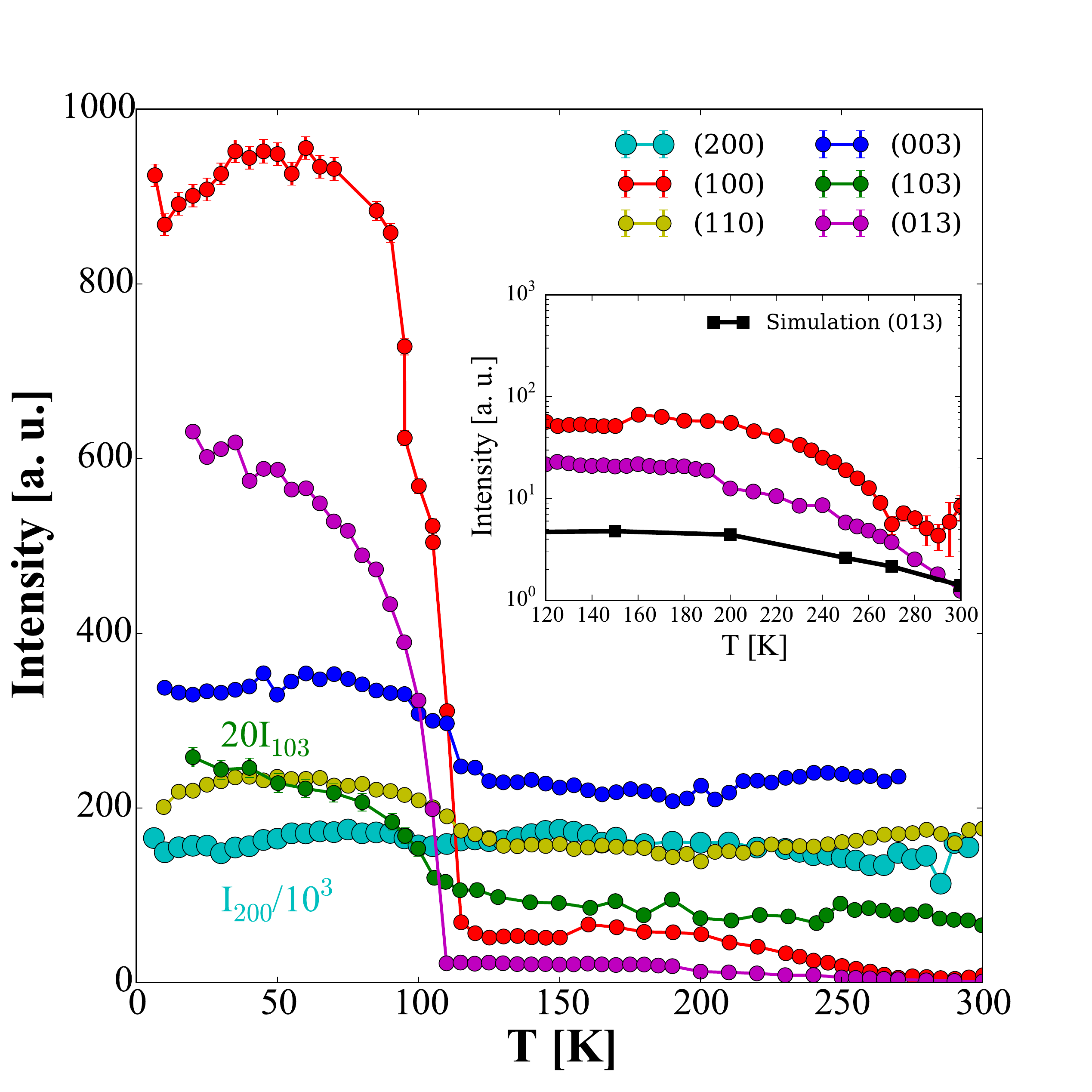}
	\caption{(Colour online) Temperature dependence of several reflections in the insulating region. (inset) Magnetic reflections associated to $m_b$ highlighted in logarithmic scale, with a comparison (see main text for details) to FDMNES simulations for the (013). Reflections are measured on multiple samples using an area detector, without polarisation analysis. Reflections (200), (100), (110) and (003) were measured at 2.967 keV, whereas (013) and (103) were measured at 2.838 keV.}
	\label{fig2}
\end{figure}

\begin{figure}[ht]
	\centering
	\includegraphics[width=0.4\textwidth]{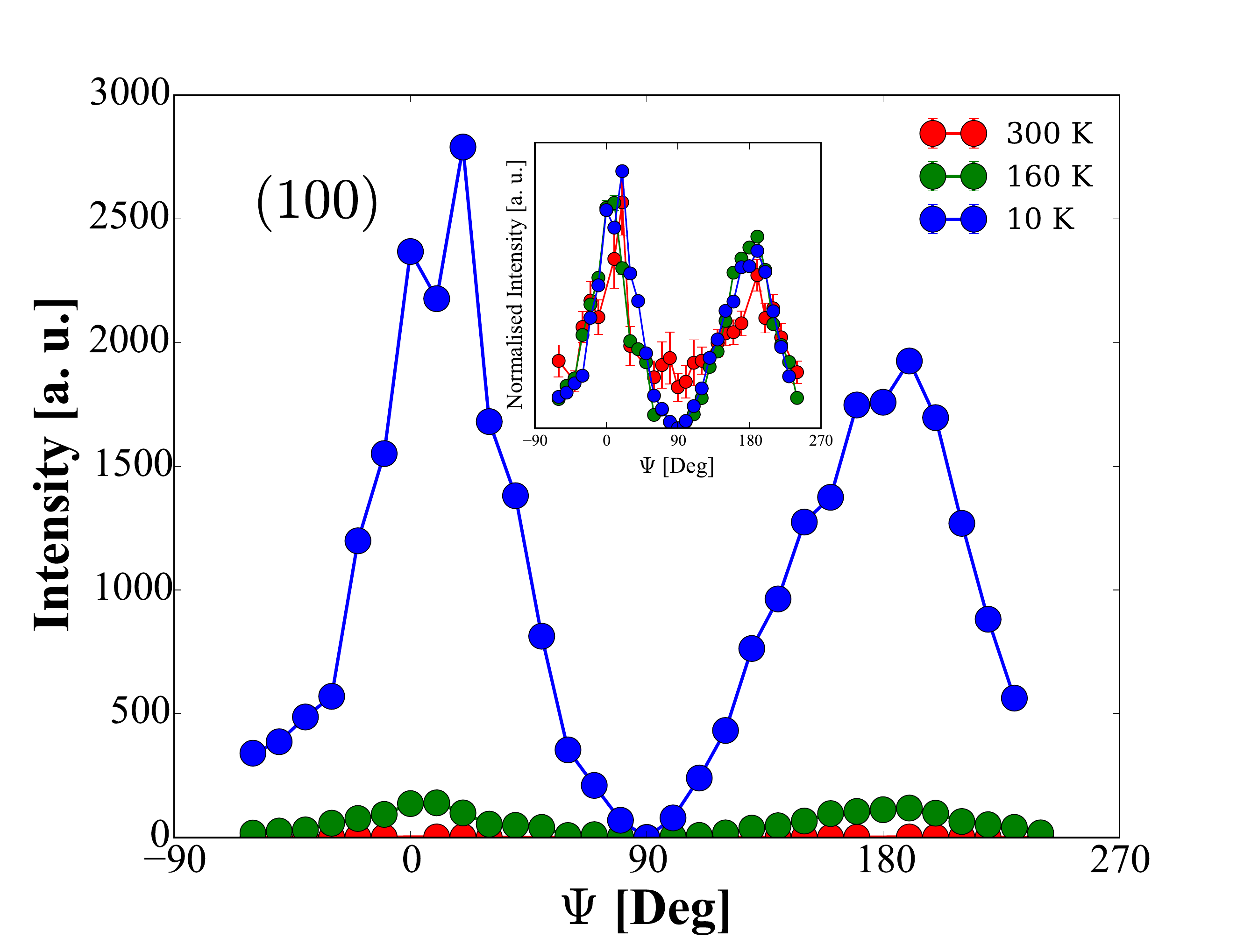} 
	\caption{(Colour online) Variation in intensity of (100) resonant reflections with azimuthal angle, well described by Eq. \ref{eq2c} at all temperatures.} \label{fig:xrayazi100}
\end{figure}

Concerning the orbital behaviour at $T_{OO}$, we report in Fig. \ref{fig2} the temperature dependence of several reflections belonging to each of the above three classes. In particular, we can see that, of all reflections, only the (100) and the (013) show some variations in intensity around $T_{OO} \approx$ 260 K. The (100) reflection was already studied previously in Ref. [\onlinecite{Zegkinoglou2005}] and our own measurements are consistent with it. Shown here in logarithmic scale (see inset of Fig. \ref{fig2}), it appears as if no real transition takes place at $T_{OO}$, at least for the (013) transition, whereas the situation is less clear for the (100), in keeping with the previous results.
It should be noted that all these reflections are allowed in resonant conditions without the need for breaking any symmetry, so that the assignment of T$_{OO}$ as an orbital-order phase transition on such a basis should probably be reanalyzed. Another element in this direction is that, from Fig. \ref{fig:xrayazi100}, even the reflection (100) does not show any deviation in its azimuthal dependence from the expected azimuthal behaviour of the anisotropic tensor of susceptibility (ATS scattering), in keeping with Eq. \ref{eq2c}. We remind that ATS scattering is determined by the crystal space group (as in Eq. (\ref{eq1})). 

In this view we might wonder if the continuous contraction of the $c$-axis, reported in Fig. \ref{fig:xrd}, as well as the temperature dependence of the Ru-O$_2$ bond, with decreasing temperature might lead to a bigger crystal field along the $c$-axis, leading in turn to a corresponding increase in the population of the $d_{xy}$ orbital and therefore an increase in the empty $d_{xz}$ and $d_{yz}$ orbitals. As REXS at both (100) and (013) reflections mainly probes empty $d_{xz}$ and $d_{yz}$ orbitals, as shown in Table \ref{tab:ref_sym}, this might explain at least the (013) signal, without the need to invoke an extra orbital-order origin for this temperature dependence. 
We tried therefore to simulate the increase in the (013) signal using FDMNES\cite{Bunau2009} with the temperature-dependent experimental lattice parameters and the refined atomic positions given in Table \ref{tab:xrd}. The result is shown in the inset to Fig. \ref{fig2}. Though the trend with temperature is correct, the theoretical calculation produces a smaller increase compared to the values experimentally measured. Such findings might be interpreted in terms of an extra orbital ordering due to the electronic correlations (i.e., beyond the one induced by the crystal-field variations), with the same symmetry as the local crystal field, that cooperate to produce the intensity increase. We can state something more precise about this extra orbital ordering by reminding that both the (013) and (100) reflections are sensitive to the specific linear combination (limiting to the ($d_{xz}$, $d_{yz}$) subspace, see Eq. \ref{Qxz}): $0.49 d_{xz} - 0.78 d_{yz}$. We might therefore conclude that the orbital that reduces its population in the ($d_{xz}$, $d_{yz}$) subspace is the one that respects the above phase condition and not the orthogonal one. Unfortunately, the FDMNES calculation does not reproduce such a phase condition and for this reason misses a big part of the spectral weight. We turn therefore in the next subsection to a model hamiltonian that can at least well describe the magnetic Ru-Ru correlations at T$_N$, explaining in this way the magnetic canting experimentally found in Subsection IV.A.      

\subsection{Theoretical model of the magnetic canting below T$_N$.}

In this Subsection we investigate the magnetic anisotropy in the antiferromagnetic phase of \cro{}. We demonstrate that the nearest neighbour Ru-Ru spin correlations are antiferromagnetic with a dominant in-plane easy axis but also a non-vanishing out-of-plane component. In particular, in the electronic regime that is relevant for the \cro{}, one finds that there is always an out-of-plane magnetic component which is related to the interplay between the crystal field potential, associated with a flat octahedral configuration, and the atomic spin-orbit coupling. 
The analysis is performed by considering both the low-energy description, after integrating out the high-energy charge degrees of freedom, and by solving a model Hamiltonian for a Ru-O-Ru cluster that is able to capture all the electronic processes that contribute to set the magnetic exchange. 

Concerning the Ru-O-Ru cluster, we employ a model Hamiltonian for the bands close to the Fermi level for the itinerant electrons within the ruthenium-oxygen plane which includes the interaction terms at the Ru and O sites and the kinetic part for the Ru-O connectivity.
The local ruthenium Hamiltonian $H_{\mathrm{loc}}$,\cite{Cuoco2006a,Cuoco2006b} consists of the complete Coulomb interaction for the t$_{2g}$ electrons, the spin-orbit coupling, and the tetragonal crystal field potential. The on-site Coulomb, spin-orbit and crystal field contributions are given by:
\begin{eqnarray}
H_{\text{el-el}}(i) 
&=&
U\sum n_{i\alpha \uparrow }n_{i\alpha \downarrow}
-2J_{\mathrm{H}}\sum\limits_{\alpha <\beta }\bf{S}_{i\alpha }\cdot 
\bf{S}_{i\beta }
+\nonumber \\&&+
\left(U-\frac{5 J_{\mathrm{H}}}{2}\right)\sum\limits_{\alpha <\beta }n_{i\alpha }n_{i\beta }
+\nonumber \\&&+J_{H}\sum\limits_{\alpha <\beta }d_{i\alpha \uparrow }^{\dagger }d_{i\alpha
\downarrow }^{\dagger }d_{i\beta \uparrow }d_{i\beta \downarrow }, \nonumber \\
H_{\mathrm{SOC}}(i) &=&\lambda \sum\limits_{\alpha ,\sigma }\sum_{\beta ,\sigma
^{^{\prime }}} d_{i\alpha \sigma }^{\dagger } (\bf{l}_{\alpha\beta }\cdot {\bf{s}_{\sigma \sigma ^{^{\prime }}}})
d_{i\beta
\sigma ^{^{\prime }}} ,\nonumber \\
H_{\mathrm{cf}}(i) &=&\varepsilon_{xy} n_{i,xy}+\varepsilon_{z}\left(
n_{i,xz}+n_{i,yz}\right), \nonumber \\
H_{\mathrm{loc}}(i) &=&H_{H_{\text{el-el}}}(i)+H_{\mathrm{SOC}}(i)+H_{\mathrm{cf}}(i)\, \nonumber,
\end{eqnarray}

\noindent where $i$ labels the site and $\alpha,\beta $ are indices running over the three orbitals in the t$_{2g}$ sector, i.e. $\alpha,\beta \in \{d_{xy},d_{xz},d_{yz}\}$, and $d_{i\alpha \sigma }^{\dagger }$ is the creation operator of an electron with spin $\sigma $ at the site $i$ in the orbital $\alpha$. The interaction is parametrized by the intra-orbital Coulomb interaction $U$ and the Hund's coupling  $J_{\mathrm{H}}$. The strength of the tetragonal distortions is expressed by the amplitude $\delta$, with $\delta=(\varepsilon_{xy}-\varepsilon_z)$. 
Furthermore, we consider the ruthenium-oxygen hopping, which includes all the allowed symmetry terms according to the Slater-Koster rules \cite{Slater} for a given bond connecting a ruthenium to an oxygen atom along, say, the $x$-direction. Here, we allow for the relative rotation of the octahedra assuming that the Ru-O-Ru bond can form an angle $\theta=(180^{\circ}-\phi)$. The case with $\phi=0$ corresponds to the tetragonal undistorted bond, while a non-vanishing value of $\phi$ arises when the RuO$_6$ octahedra are rotated of the corresponding angle around the $c$-axis. 
The angular dependence of the Ru-O-Ru bond and $\phi = \sin^{-1}\left( \vec{RuO1}\cdot\left(\vec{a}+\vec{b}\right) \right)$ are given in Table \ref{tab:xrd} and Figure \ref{fig:xrd}(b). 

Since we are interested in the spin-spin correlations of the nearest neighbour Ru-Ru, it is useful to introduce the total spin operator at each Ru site as ${\bf{S}}_{\bf{i}}={\bf{S}}_{{\bf{i}}xy}+{\bf{S}}_{{\bf{i}}xz}+{\bf{S}}_{{\bf{i}}yz}$, with ${\bf{i}}=1,2$ labelling the two Ru sites. To evaluate the model, we determine the ground state for the Ru-O-Ru cluster and the ensuing Ru-Ru spin correlations for the planar ${\bf{S}}_{{\bf{i}}ab}$ and out-of-plane components ${\bf{S}}_{{\bf{i}}c}$, respectively. 
For the Ca$_2$RuO$_4$ system, since the octahedra become flat below the structural transition, $\delta$ is negative and, according to first principle calculations or estimates employed to reproduce the resonant inelastic x-ray \cite{Das2018} and the neutron scattering spectra,\cite{Jain2017} its amplitude is in the range $\sim$ [200-300] meV.
To evaluate the model and the magnetic properties of the ground state, material speciﬁc values $\lambda=0.075$ eV, $U=2$ eV, and $J_{H}$ in the range [0.35, 0.5] eV,\cite{Mizokawa2001, Veenstra2014} are used. Similar values for $\delta$, $U$ and $J_H$ have been used for calculations \cite{Sutter2017} of electronic spectra in Ca$_2$RuO$_4$ and in other class of ruthenates,\cite{Granata2016,Forte2010,Malvestuto2013} with the ratio $g=\delta/(2 \lambda)$ in the range $\sim$[1.5,2] when modelling the spin excitations observed by neutron and Raman scattering.\cite{Jain2017,Souliou2017}
For the hopping amplitudes, we consider a representative set of electronic parameters for the Ru-O-Ru cluster that is consistent with typical amplitudes obtained from first-principles calculations for the class of ruthenates.\cite{Fang2004,Gorelov2010,Das2018}

\begin{figure}[t]
\includegraphics[width=0.95\columnwidth]{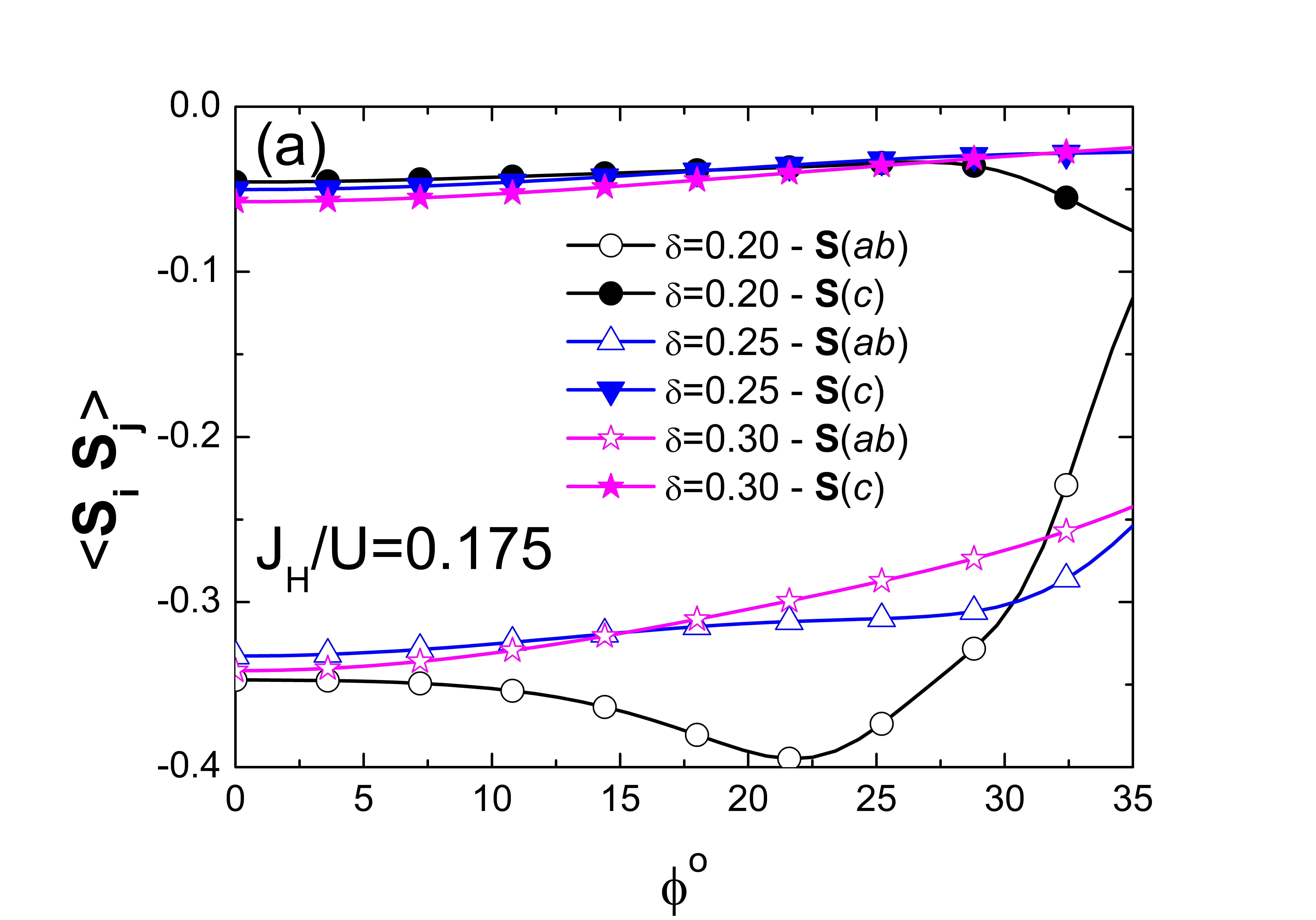}
\includegraphics[width=0.95\columnwidth]{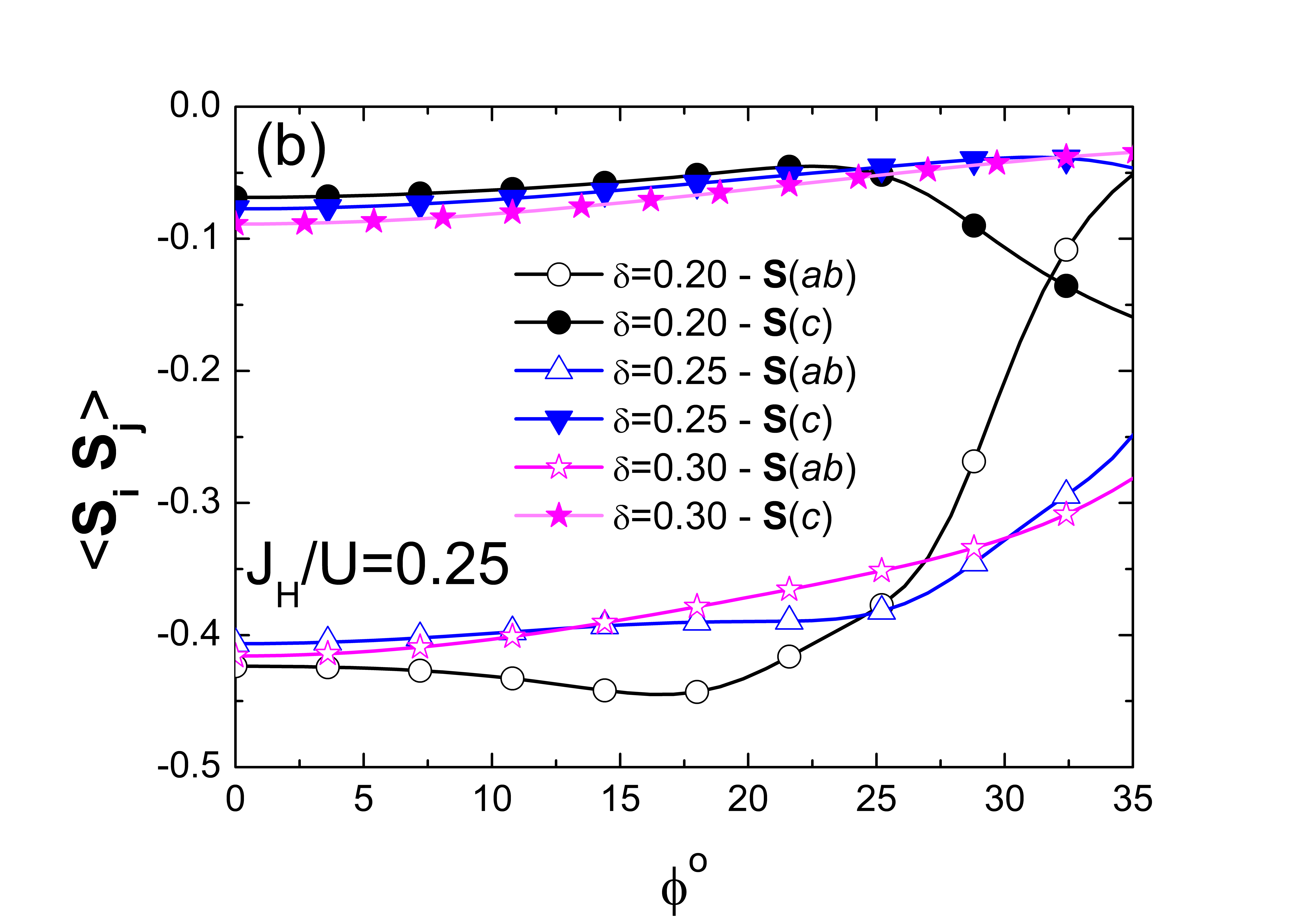}
\caption{(Colour  online). In plane (${\bf S}_{ab}$) and out-of-plane (${\bf S}_{c}$) Ru-Ru spin correlations evaluated in the ground state of the Ru-O-Ru cluster as a function of the Ru-O-Ru bond angle $\phi$ at different values of the crystal field potential for two ratio of the Hund and Coulomb interaction, i.e. (a) $J_{H}/U=0.175$, and (b)  $J_{H}/U=0.25$. }
\label{fig:fig1}
\end{figure}

Due to the competing spin-orbit and crystal field potential, the spin correlations are expected to be anisotropic. Indeed, in the $d^4$ ground-state configuration for the ruthenium, with two electrons occupying the $xy$ orbital stabilized by the compressive tetragonal distortion, the remaining orbitals $yz$ and $zx$ are mostly half-filled due to Coulomb interaction, and thus the spin-orbit $l_x$ and $l_y$ components are the dominant ones to set the in-plane magnetic exchange. This is a general behaviour which is obtained when both varying the angle of relative rotation of the octahedra around the $c$-axis as well as the crystal field splitting associated to compressed octahedra. In Fig. \ref{fig:fig1} we report the in-plane and out-of-plane Ru-Ru spin correlations evaluated in the ground state of the Ru-O-Ru cluster for different octahedral configurations and electronic regimes in terms of Coulomb and Hund couplings. As one can notice, the change of the Ru-O-Ru bond angle from the tetragonal ($\phi=0$) to the distorted one ($\phi \neq0$) generally tends to decrease the in-plane and out-of-plane Ru-Ru spin correlations except for the case of small compression of the octahedra. When the ratio $g$ gets close to one, e.g. with $\delta=0.2$ eV, the evolution of the Ru-Ru spin-correlations is non-monotonous and for amplitude of the angles $\phi$ above $\sim 25^{\circ}$ the in-plane and out-of-plane components tend to a spin-isotropic limit. Such behaviour is slightly modified by a change in the ratio $J_{H}/U$ in the direction of shifting the isotropic regime to higher Ru-O-Ru bond bending when reducing $J_{H}/U$. 
Although the regime of Ru-O-Ru bond angles larger than $\sim 25^{\circ}$  is not directly relevant for the Ca$_2$RuO$_4$ compound, the changeover of the spin correlations emphasizes the interplay of the compression and rotation of the RuO$_6$ octahedra in setting the anisotropy of the antiferromagnet.   
Finally, we find that the reduction of the Hund's coupling generally tends to decrease the amplitude of the nearest neighbours spin correlations (see Fig. \ref{fig:fig1}).      

In order to further understand the origin of the magnetic anisotropy, it is useful to evaluate the low-energy processes that couple the spin and orbital moments of the Ru atoms. In the $d^4$ configuration at the Ru site with $\delta/(2 \lambda)>0.5$ the local lowest energy-states are well separated by the rest of the spectrum \cite{Jain2017,Das2018} and the spin-orbital  magnetism can be described by an effective pseudospin $T=1$ Hamiltonian \cite{Jain2017}:
\begin{eqnarray}
H_{eff}&=&\sum_{\langle ij \rangle} \{ J_{xy} [T_{xi}\cdot T_{xj}+ T_{yi}\cdot T_{yj}] + J_{z} T_{zi}\cdot T_{zj} \} + \nonumber \\&& E_z \sum_{i} T^2_{zi} + \sum_{\langle ij \rangle} [J_{1x} T_{yi}\cdot T_{zj}+ J_{1y} T_{zi}\cdot T_{xj}+ \nonumber \\&&  J_{1z} T_{xi}\cdot T_{yj} + {\mathrm h.c.}] \,.
\end{eqnarray}
with $E_z$ being greater than $J_{xy}$,$J_{z}$ and the anisotropic couplings $J_{1x},J_{1y},J_{1z}$ and $J_{1\alpha} < J_{xy}$,$J_z$.

The local basis $\{|0\rangle,|+1\rangle,|-1\rangle\}$ for the $T=1$ pseudospin are expressed in terms of the original spin and orbital eigenstates $|S_z,L_z\rangle$ of the corresponding operators with angular momentum one as \cite{Jain2017}:
\begin{eqnarray}
|0 \rangle &=&\sin(\theta_0) \frac{1}{\sqrt{2}} \left[|1,-1\rangle+|-1,1\rangle\right]- \cos(\theta_0) |0,0\rangle\nonumber \\
|+1\rangle&=& \cos(\theta_1) |1,0\rangle - \sin(\theta_1) |0,1\rangle \nonumber \\
|-1\rangle&=& -\cos(\theta_1) |-1,0\rangle + \sin(\theta_1) |0,-1\rangle \nonumber
\end{eqnarray}
with $\tan(\theta_1)=\frac{1}{1+\sqrt{1+g^2}}$ and $\tan(\theta_0)= \sqrt{1+\beta^2}-\beta$, with $\beta=\frac{1}{\sqrt{2}}(g-\frac{1}{2})$.

Since the anisotropic splitting $E_z$ is the largest energy scale, then the ground state is mainly due to pseudospin lying the $xy$ plane. For such configurations, one can demonstrate that the spin correlations are also made of only in-plane components. However, the terms proportional to $J_{1x}$ and $J_{1y}$ in $H_{eff}$ induce out-of-plane pseudospin correlations which in turn yield antiferromagnetic spin correlations consistently with the analysis performed on the Ru-O-Ru cluster. In particular, considering the ground state configuration of the model Hamiltonian $H_{eff}$, with constraints consistent with the second-order perturbation theory, one can show that the nearest neighbour in-plane Ru-Ru spin correlations ${\bf S}({\it ab})_{ij}$ are antiferromagnetic and proportional to $\sim -(\cos \theta_1 \cos \theta_0 +\frac{1}{\sqrt{2}} \sin \theta_1 \sin \theta_0)^2$, while the out-of-plane spin correlations ${\bf S}({\it c})_{ij}$ scales as $\sim -(\cos \theta_1 \cos \theta_0)^2$. As one can notice, the sign of the spin-correlations is always negative, independently of the amplitudes of the electronic parameters.
It is important to point out that the analysis performed on the Ru-O-Ru bond applies also when considering a planar cluster with larger size. Since there are no frustrating interactions, no qualitative variations in the character of the spin correlations and of the magnetic anisotropy are expected.

Finally,\footnote{We thank an anonymous Referee for suggesting such a comparison} we would like to highlight the different physical mechanisms acting in the cases of Ca$_2$RuO$_4$ and of the superconducting cuprate parent compound La$_2$CuO$_4$. The latter is also characterized by a tilting along the c-axis of the in-plane magnetic moment.\cite{benfatto} In spite of the apparent analogy, however, the two cases are different for the following three reasons: 
1) The tilting along the c-axis of the magnetic moment in La$_2$CuO$_4$ leads to a ferromagnetic component along the c-axis, as measured by XMCD.\cite{DeLuca} This is not the case for Ca$_2$RuO$_4$, as the c-component of the magnetic moment is compensated at each ab-plane (if it is $+m_c$ at Ru$_1$, then it is $-m_c$ at Ru$_4$, see Sec. II).
2) Copper magnetism in La$_2$CuO$_4$ mainly takes place through the spin of the single d$_{x^2-y^2}$ hole and the mechanism behind the canting is a pure spin Dzjaloshinskii-Moriya effect.\cite{benfatto} In our case, as explained above, orbital degrees of freedom cannot be neglected, and the coupled spin-orbital Hamiltonian, H$_{\rm eff}$, is necessary to explain the sign of the coupling through the entanglement of the orbital components, a degree of freedom which is absent in the La$_2$CuO$_4$ case. 
3) As described in Section II, in Ca$_2$RuO$_4$ the canting is allowed by the point symmetry ($\overline{1}$), whereas this is not the case for the Cu point symmetry in La$_2$CuO$_4$ at room temperature, which is 2/m (space group N. 64, Cmca\cite{vaknin}).

\section{Conclusions}
An accurate description of the magnetic and orbital pattern is crucial in a complex system as 
\cro{} presenting such an intricate interplay of the orbital and magnetic degrees of freedom. 
In this paper we comprehensively explored the magnetic and orbital ordering in an extremely pure sample
of \cro{}, clarifying the relationship between each accessible observable and the scattering condition in the dipolar approximation.  
Our analysis indicates that the magnetic moment is not confined along the $b$ axis as previously suggested but a fraction of about one tenth of the moment points along the $c$-axis. This antiferromagnetic canting is symmetry allowed and can be quantitatively understood considering the theoretical model introduced in Subsection IV.C: antiferromagnetic spin correlations between nearest neighbour Ru-Ru ions, combined with an interplay between the crystal field potential and spin-orbit coupling, lead to an always out-of-plane magnetic component as well as to a dominant in-plane easy axis.
It was not necessary to invoke any change in the orbital arrangement through T$_{AF}$ to model the experimental findings.
It is interesting to remark that the theoretical findings of in-plane antiferromagnetic correlations, combined with the constraint of a two-fold symmetry ${\hat{A}}_{2}$ around the a-axis for the two nearest-neighbor in-plane Ru sites, necessarily implies that the $m_a$ component of the magnetic moment along the a-axis is zero, as experimentally found. In fact, as shown in Section II, the two-fold symmetry axis ${\hat{A}}_{2}$ leads to antiferromagnetic correlations for the $m_b$ component and ferromagnetic correlations for the $m_a$ component, the latter being excluded by the model Hamiltonian. This leads to a zero value for $m_a$.

As detailed in Subsection IV.B, no symmetry breaking was measured near T$_{OO}$, related to the signal increase in the (013) and (100) reflections. This aspect, together with the different behaviours of two reflections (as seen from the logarithmic scale) probing the same tensors across this ``transition'', makes the origin of this signal unclear. Even if it would be possible to explain qualitatively the signal as purely due to the contraction of the octahedra along the $c$-axis, the variation in amplitude obtained from our simulations quantitatively disagrees with the experimental observation, suggesting that probably a complex phase relation within the ($d_{xz}$, $d_{yz}$) manifold is possibly at the origin of the experimental observation.

\begin{acknowledgments}
We acknowledge Diamond Light Source for time on Beamline I16 under Proposals 12609, 14815 and 17569.  Images of atomic structures were created using the computer program VESTA, by Koichi Momma and Fujio Izumi\cite{Momma2011}. 
\end{acknowledgments}

\appendix
\section {Structure factor and orbital rotation.}

In the resonant regime the elastic scattering amplitude, $A({\vec q},\omega)$, can be  written in terms of the atomic scattering factors (ASF), $f_j(\omega)$, of the atoms at positions $\vec{\rho}_j$ in the unit cell as:

\begin{equation}
A(\vec{q},\omega) = \Sigma _j e^{i\vec{q} \cdot \vec{\rho_j}} f_j(\omega)
\label{bragg}
\end{equation}

\noindent where $\hbar \vec{q}$ is the momentum transfer in the scattering process, $\hbar \omega$ the incoming and outgoing photon energy and the sum is over the four atoms in the orthorhombic primitive cell. The ASF is a second-order process in the electron-radiation interaction, whose explicit expression, for core resonances in the dipole-dipole approximation and linearly scattered polarization, reads, in atomic units:

\begin{equation}
f_j(\omega) = (\hbar \omega)^2 \sum _n \frac{\langle \Psi_0^{(j)}|\vec{\epsilon}^s\cdot\vec{r}|\Psi_n \rangle \langle \Psi_n|\vec{\epsilon}^i\cdot\vec{r}|\Psi_0^{(j)}\rangle}{\hbar \omega - (E_n-E_0) -i\Gamma_n}
\label{asf}
\end{equation}

\noindent where $|\Psi_0^{(j)}\rangle$ is the ground state, with the origin taken on the $j$-th scattering atom, and  $E_0$ its energy; the sum is over all the excited states $|\Psi_n \rangle$, with corresponding energies $E_n$. Finally $\Gamma_n$ is a damping term that takes into account the core-hole and the finite life-time of the excited states $|\Psi_n \rangle$. The indices $i, s$ refer to the incident (scattered) properties of the polarization $\vec{\epsilon}$ and $r$ is the coordinate of the electron in the reference frame of the scattering atom. As the matrix element in Eq. (\ref{asf}) are independent of the photon polarization, the latter can be factorised, so as to obtain:

\begin{equation}
f_j(\omega)= \Sigma_{\alpha\beta} \epsilon^s_{\alpha}\epsilon^i_{\beta}f_j^{\alpha\beta}(\omega)
\label{asffinal}
\end{equation}

\noindent where we introduced the matter-tensor $f_j^{\alpha\beta}(\omega) \equiv \langle \Psi_0^{(j)}| \sum _n \frac{{r}_{\alpha}|\Psi_n \rangle \langle \Psi_n|{r}_{\beta}}{\hbar \omega - (E_n-E_0) -i\Gamma_n} |\Psi_0^{(j)}\rangle$ that couples scalarly to the polarization tensor, $\epsilon^s_{\alpha}\epsilon^i_{\beta}$
(here $\alpha$, $\beta$ are cartesian coordinates). The matter tensors at the four equivalent atomic sites $j$ belonging to the $4a$ Wyckoff positions are related one another by the symmetry operations of {\it Pbca}, as listed in the main text. 

The tensor $f_j^{\alpha\beta}$ in the dipole-dipole channel is composed of three irreducible parts\cite{sdm2012}: a scalar part, not contributing to the analyzed Bragg-forbidden reflections; the antisymmetric part (an axial-vector), proportional to the components of the magnetic moment, $m_a$, $m_b$, $m_c$ along the three crystallographic axes; the traceless symmetric part, five components ($Q_{ab}$, $Q_{ac}$, $Q_{bc}$, $Q_{2c^2-a^2-b^2}$, $Q_{a^2-b^2}$) related to the anisotropy at the Ru site, proportional, at the $t_{2g}$ energies, to the degree of OO and, at the $e_g$ energies, to the crystal field of the surrounding oxygen octahedron.

From Eq. \ref{eq1}, we get the following conclusions, summarized in Table \ref{tab:ref_sym}: 
{\it 1) -} Reflections with $k+l=$ odd and $h+l=$ even. In this case, e.g., at the (103): $F_{103} = \left(1 + {\hat{C}}_{2}\right) \left(1 - {\hat{B}}_{2} \right) f_1^{\alpha\beta}$, and only the $m_c$ component can be detected, as it changes sign under ${\hat{B}}_{2}$  and does not under ${\hat{C}}_{2}$. For the same reason, the only non-zero component of the electric quadrupole, measuring at L$_{2,3}$-edges the orbital anisotropy, is $Q_{ab}$. 

{\it 2) -} Reflections with $k+l=$ even and $h+l=$ odd. In this case, e.g., at the (013): $F_{013} = \left(1 - {\hat{C}}_{2}\right) \left(1 + {\hat{B}}_{2} \right) f_1^{\alpha\beta}$. The only non-zero components are $m_b$ and $Q_{ac}$, that change sign under ${\hat{C}}_{2}$ and does not under ${\hat{B}}_{2}$. 

{\it 3) -} Reflections with $k+l=$ odd and $h+l=$ odd. In this case, e.g., at the (110): $F_{110} = \left(1 - {\hat{C}}_{2}\right) \left(1 - {\hat{B}}_{2} \right) f_1^{\alpha\beta}$. The only non-zero components in this case are $m_a$ and $Q_{bc}$, that change sign under both ${\hat{C}}_{2}$ and ${\hat{B}}_{2}$.

Interestingly, we can perform the same analysis of the main text for {\it Pbc'a'} magnetic space group (also known in the literature as B-centred magnetic structure), corresponding to the 1\% Ti-doped compound \cite{Kunkemoller2017}. Clearly, the only difference appears for magnetic reflections, due to the action of the ${\hat{T}}$ operator that reverses the magnetic signal associated to ${\hat{C}}_{2}$ and ${\hat{B}}_{2}$ operations. We get in this case:

\begin{align}
F_{hkl} &= f_1 + (-)^{h+l} f_2 + (-)^{k+l} f_3 + (-)^{h+k} f_4 \nonumber \\ 
&= \left(1 + (-)^{h+l}{\hat{T}}{\hat{C}}_{2}\right) \left(1 + (-)^{k+l}{\hat{T}}{\hat{B}}_{2} \right) f_1^{\alpha\beta}
\label{eq1B}
\end{align} 

Therefore:

{\it 1) -} Reflections with $h+l=$even and $k+l=$odd. In this case: $F^A_{103} = \left(1 + {\hat{T}}{\hat{C}}_{2}\right) \left(1 - {\hat{T}}{\hat{B}}_{2} \right) f_1^{\alpha\beta}$. The electric quadrupole, non-magnetic, behaves like the {\it Pbca} case of the main text ($Q_{ab}$). The magnetic channel, instead, is proportional to $m_b$, that changes sign under ${\hat{T}}{\hat{B}}_{2}$ and does not change sign under ${\hat{T}}{\hat{C}}_{2}$. We remark the difference with the {\it Pbca} case, where $m_c$ was detected at these reflections. This implies that a magnetic signal at the (103) can either be a consequence of an $m_c$ component in a $100 \%$ {\it Pbca} crystal, or a 'contamination' of a {\it Pbc'a'} component. It was possible however to settle unambiguously the question in favour of {\it Pbca} by performing an azimuthal scan. The {\it Pbca} case behaves like $0.44\sin\psi - 0.535$, as experimentally found, whereas the {\it Pbc'a'} case would have led to $0.44\cos\psi$. 

{\it 2) -} Reflections with $k+l=$even and $h+l=$odd. In this case: $F^A_{013} = \left(1 - {\hat{T}}{\hat{C}}_{2}\right) \left(1 + {\hat{T}}{\hat{B}}_{2} \right) f_1^{\alpha\beta}$. The electric quadrupole, non-magnetic, behaves like the {\it Pbca} case of the main text ($Q_{ac}$).
The magnetic channel, instead, is proportional to $m_c$, that changes sign under ${\hat{T}}{\hat{C}}_{2}$ and does not change sign under ${\hat{T}}{\hat{B}}_{2}$. 

{\it 3) -} Reflections with $k+l=$odd and $h+l=$odd. In this case: $F^A_{110} = \left(1 - {\hat{T}}{\hat{C}}_{2}\right) \left(1 - {\hat{T}}{\hat{B}}_{2} \right) f_1^{\alpha\beta}$. The electric quadrupole, non-magnetic, behaves like the {\it Pbca} case of the main text ($Q_{bc}$). In the magnetic channel, instead, no signal is allowed, either of the two factors being zero.

\section {Relation of REXS crystallographic tensors and $4d$ orbitals}
The electric-quadrupole expectation values derived in the main text are related to the crystallographic axes of the {\it Pbca} setting. However, in all theoretical descriptions of the system, the $4d$ orbitals involved are related to the local octahedral frame around Ru-ions. We call it the $xyz$-frame (orthonormal, see Fig. \ref{fig1}). We remark that here we do not consider the extra tilting of the apical oxygens compared to the plane rotation. 

The rotation to pass from the $abc$-frame to the $xyz$-frame is identified by the three Euler angles $\alpha$, $\beta$ and $\gamma$ given by:

\begin{itemize}
\item $\alpha=-\pi/4$ is the rotation angle around the $z$ axis so as to bring the $y$ axis along the line of node of the two frames ($xyz$  and $abc$). Notice that this rotation can be $\pm \pi/4$ according to what is the original choice of $x$ and $y$ in the local frame (we have chosen the b-axis pointing in the positive xy-directions, see Fig. \ref{fig1}). The two choices are not equivalent, the system being orthorhombic and not tetragonal.
\item $\beta \sim -11\pi/180$ is the rotation angle around the line of nodes of the two frames that leads $z$ to coincide with $c$. It corresponds to the angle called $\theta$ in [\onlinecite{Braden1998}], the angle of buckling of the oxygen plane with respect to the $\vec{a}-\vec{b}$ plane of the {\it Pbca} setting. Notice that, as stated above, we consider at this stage just one rigid rotation of the octahedra, thereby neglecting the small difference between the rotation of the 4 planar oxygen and of the 2 apical oxygen ions. We also notice that the angle $\beta$ is temperature dependent, going from $\sim -9\pi/180$ to $\sim -13\pi/180$.
\item $\gamma \simeq -11.88\pi/180$ is the final rotation around the $c$ axis. It corresponds to the angle $\phi$ in Ref. [\onlinecite{Braden1998}], which is temperature independent.
\end{itemize}

Notice that all rotations are negative, because clockwise.

All spherical harmonics rotate through Wigner's matrices\cite{k1988quantum}:

\begin{eqnarray}
Y_{lm}(\theta,\phi)= \sum_{m'} Y_{lm'}(\theta',\phi') e^{-im'\alpha} d^l_{m',m}(\beta) e^{-im\gamma}
\label{Ygenrot}
\end{eqnarray}

where $d^l_{m',m}(\beta)$ are the reduced Wigner matrices.\cite{k1988quantum}

In this way, reminding that the electric-quadrupole REXS expectation values $Q$ behave like rank-two spherical tensors and in the hypothesis (see below) that the only available density of states with $l=2$ character originates from $4d$ Ru orbitals, we obtain the following expressions in terms of the local $4d$ orbitals of Ru ions:

\small
\begin{align} \label{Qyz}
& Q_{bc}= d_{3z^2-r^2}\sqrt{3} \sin\beta\cos\beta \sin\gamma  \\
&+d_{yz} \left(\cos\alpha \cos\beta \cos\gamma - \sin\alpha (2\cos^2\beta-1)\sin\gamma \right) \nonumber \\
&-d_{xz} \left(\sin\alpha \cos\beta \cos\gamma + \cos\alpha (2\cos^2\beta-1)\sin\gamma \right) \nonumber \\
&+d_{xy} \left(\cos(2\alpha) \sin\beta \cos\gamma - \sin(2\alpha)\sin\beta\cos\beta \sin\gamma \right) \nonumber \\
&-d_{x^2-y^2} \left(\cos(2\alpha) \sin\beta \cos\beta \sin\gamma + \sin(2\alpha)\sin\beta\cos\gamma \right) \nonumber \\
&\simeq 0.54 d_{yz} +0.81 d_{xz} -0.04 d_{xy} +0.20 d_{x^2-y^2} +0.07 d_{3z^2-r^2} \nonumber
\end{align}
\normalsize

\small
\begin{align} \label{Qxz}
& Q_{ac}= d_{3z^2-r^2} \sqrt{3} \sin\beta\cos\beta \cos\gamma \\
&+d_{yz} \left(\sin\alpha (2\cos^2\beta-1) \cos\gamma + \cos\alpha \cos\beta \sin\gamma \right) \nonumber \\
&+d_{xz} \left(\cos\alpha (2\cos^2\beta-1) \cos\gamma - \sin\alpha \cos\beta \sin\gamma \right) \nonumber \\
&+d_{xy}\left(\cos(2\alpha) \sin\beta \sin\gamma - \sin(2\alpha)\sin\beta\cos\beta \cos\gamma \right) \nonumber \\
&+d_{x^2-y^2} \left(\cos(2\alpha) \sin\beta \cos\beta \cos\gamma - \sin(2\alpha)\sin\beta\sin\gamma \right) \nonumber \\
&\simeq -0.78 d_{yz} +0.49 d_{xz} +0.20 d_{xy} -0.04 d_{x^2-y^2} +0.35 d_{3z^2-r^2}  \nonumber
\end{align}
\normalsize

\small
\begin{align} \label{Qxy}
&Q_{ab}=d_{3z^2-r^2}\frac{\sqrt{3}}{2} \sin^2\beta \sin(2\gamma) \\
&+d_{yz} \left(\sin\alpha\sin\beta\cos\beta\sin(2\gamma) - \cos\alpha \sin\beta \cos(2\gamma) \right) \nonumber \\
&+d_{xz} \left(\cos\alpha\sin\beta\cos\beta\sin(2\gamma) + \sin\alpha \sin\beta \cos(2\gamma) \right) \nonumber \\
&+d_{xy} \left(\cos(2\alpha) \cos\beta \cos(2\gamma) - \sin(2\alpha) \frac{1+\cos^2\beta}{2} \sin(2\gamma) \right) \nonumber \\
&-d_{x^2-y^2} \left(\cos(2\alpha) \frac{1+\cos^2\beta}{2} \sin(2\gamma) + \sin(2\alpha)\cos\beta \cos(2\gamma) \right) \nonumber \\
&\simeq +0.07 d_{yz} +0.177 d_{xz} -0.40 d_{xy}  -0.90 d_{x^2-y^2} -0.018 d_{3z^2-r^2} \nonumber
\end{align}
\normalsize

Concerning the above hypothesis of equating the quadrupolar components to Ru $4d$ orbitals, we remind that it corresponds to neglecting possible DOS projections of ligand O $2p$ states with $l=2$ character on the central Ru.
From the previous expressions, it is possible to obtain the values of the intensities reported in Table \ref{tab:ref_sym}:
 
\small
\begin{align}
&|Q_{ac}|^2 \rightarrow 0.61 |d_{yz}|^2+ 0.24 |d_{xz}|^2 +0.04 |d_{xy}|^2 +0.12 |d_{3z^2-r^2}|^2 \nonumber \\
&|Q_{bc}|^2 \rightarrow 0.29 |d_{yz}|^2+ 0.66 |d_{xz}|^2 +0.04 |d_{x^2-y^2}|^2 +0.01 |d_{3z^2-r^2}|^2 \nonumber \\
&|Q_{ab}|^2 \rightarrow 0.005 |d_{yz}|^2+ 0.03 |d_{xz}|^2 +0.16 |d_{xy}|^2 +0.81 |d_{x^2-y^2}|^2 
\label{eq2bis}
\end{align} 
\normalsize

Unfortunately, the absence of any interference between magnetic and orbital OP makes all possible reflections insensitive to the difference between $d_1 = -(d_{xz}+id_{yz})/\sqrt{2}$ and $d_{-1} = -(d_{xz}-id_{yz})/\sqrt{2}$, which would have been an interesting parameter to probe, in the light of the possible coupling of orbital and spin magnetic moments. In fact, the previous relations can also be written as:  

\small
\begin{align}
&|Q_{ac}|^2 \rightarrow 0.85 (|d_{1}|^2+ |d_{-1}|^2) +0.04 |d_{xy}|^2 +0.12 |d_{3z^2-r^2}|^2 \nonumber \\
&|Q_{bc}|^2 \rightarrow 0.95 (|d_{1}|^2+ |d_{-1}|^2) +0.04 |d_{x^2-y^2}|^2 +0.01 |d_{3z^2-r^2}|^2 \nonumber \\
&|Q_{ab}|^2 \rightarrow 0.04 (|d_{1}|^2+ |d_{-1}|^2) +0.04 |d_{xy}|^2 +0.92 |d_{x^2-y^2}|^2 
\label{eq2ter}
\end{align} 
\normalsize

However, it is still possible to follow the relative evolution of the $d_{xy}$ orbital compared to the set $(d_{xz};d_{yz})$ (or $(d_{1};d_{-1})$) within the $t_{2g}$ manifold at reflections sensitive to $|Q_{ac}|^2$ and $|Q_{bc}|^2$.

\section{Azimuth scan of selected reflections: disentangling magnetic and orbital OPs}

All azimuth scans are referred to $\psi=0$ when the scattering plane contains the b-crystallographic axis and the projection of the scattered wave-vector $\vec{k}'$ along this axis is positive.

In the case of the (103) reflection, the vector $\vec{Q}=(103)$ makes an angle $\beta \simeq 36^{\circ}$ with the c-crystallographic axis. In the reference frame where $\vec{Q}$ is the $z$-axis and the $x$-axis is along the b-crystallographic axis, the polarization vectors can be written as: $\vec{\epsilon}_{\sigma}=(\sin\psi,\cos\psi,0)$ and $\vec{\epsilon}_{\pi_s}=(-\sin\theta_B\cos\psi, \sin\theta_B\sin\psi, \cos\theta_B)$, if we suppose to rotate the sample counterclockwise with respect to the beam around the $\vec{Q}$-vector. 
The rotation matrix to pass from the previous reference frame to the crystal reference frame is:

\begin{table}[ht]
\centering
$\begin{pmatrix}
0 &  -\cos\beta & \sin\beta  \\ 
1      &  0  & 0        \\  
0      &  \sin\beta & \cos\beta   \\
\end{pmatrix}$
\end{table}

We get therefore the following expressions for the polarization components once rotated in the $abc$-crystal frame, where Eq. \ref{eq1} applies: $\vec{\epsilon}_{\sigma}^{cr}=(-\cos\beta\cos\psi, \sin\psi, \sin\beta\cos\psi)$ and $\vec{\epsilon}_{\pi_s}^{cr}=(-\sin\theta_B\cos\beta\sin\psi +\cos\theta_B\sin\beta, -\sin\theta_B\cos\psi, \sin\theta_B\sin\beta\sin\psi +\cos\theta_B\cos\beta)$.
As in the crystallographic frame we know from Eq. (\ref{eq1}) that $Q_{xy}$ and $m_c$ couple scalarly to the polarization, it means that they have the following polarization dependence: 

\begin{align}
Q_{xy}^{\sigma\pi_s}& \propto (\epsilon_{\sigma}^{cr})_x (\epsilon_{\pi_s}^{cr})_y + (\epsilon_{\sigma}^{cr})_y (\epsilon_{\pi_s}^{cr})_x  \nonumber \\
&\simeq 0.535 \cos(2\psi) +0.44 \sin\psi \\
Q_{xy}^{\sigma\sigma_s}& \propto 2(\epsilon_{\sigma}^{cr})_x (\epsilon_{\sigma}^{cr})_y  \nonumber \\
&\simeq -0.75 \sin(2\psi) \\
m_{z}^{\sigma\pi_s}& \propto (\epsilon_{\sigma}^{cr})_x (\epsilon_{\pi_s}^{cr})_y - (\epsilon_{\sigma}^{cr})_y (\epsilon_{\pi_s}^{cr})_x  \nonumber \\
&\simeq 0.535 -0.44 \sin\psi\\
 \nonumber
\end{align}

\noindent with $\theta_B\sim 41.19^{\circ}$ at the (103) for the L$_2$-edge energy. From these amplitudes, we can get the azimuth scan of the intensity in both the $\sigma\sigma$ and $\sigma\pi_s$ channels in terms of the square of the OP $m_c$ and $Q_{ab}$ (reminding that magnetic and not magnetic quantities do not interfere in this case):
$I_{103}^{\sigma\pi_s} \propto (0.535 \cos(2\psi) +0.44 \sin\psi)^2 Q_{ab}^2 + (0.535 -0.44 \sin\psi)^2m_c^2$ and $I_{103}^{\sigma\sigma} \propto 0.56 \sin^2(2\psi) Q_{ab}^2$

We remark the four-fold dependence of the OO term, in contrast to the two-fold dependence of the magnetic moment. This is common to all the reflections, calculated below, except for on-axis reflections, where both terms have the same two-fold dependence, so that their disentanglement through the azimuth scan is no more possible. This aspect becomes clear by thinking to the appearance, of, e.g., the $d_{ac}$ orbital and the $m_b$ magnetic moment if seen from the (100) direction: both have the same two-fold dependence. We remark that Eq. (\ref{eq1}) does not allow to detect, e.g., the $d_{ac}$ orbital and the $m_b$ magnetic moment at the (001) direction, the only on-axis direction where they would appear different. Indeed, $d_{ac}$ and $m_b$ can only be seen at (2n+1,0,0) reflections; $d_{bc}$ and $m_a$ can only be seen at (0,0,2n+1) reflections and $d_{ab}$ and $m_c$ can only be seen from (0,2n+1,0) reflections. In all cases they appear with the same two-fold azimuth scan at these reflections.

We list here the azimuth dependence for all reflections given in Table \ref{tab:ref_sym}, deduced on the basis of analogous calculations as above, with the angle $\theta_B$ corresponding to the L$_2$-edge energy:  

\begin{align}
& I_{013}^{\sigma\pi_s} & \propto  (0.37 \cos(2\psi) +0.62 \cos\psi)^2 Q_{ac}^2 \nonumber \\
&                               & + (0.37 +0.62 \cos\psi)^2m_b^2 ; \nonumber \\
& I_{013}^{\sigma\sigma} & \propto 0.32 \sin^2(2\psi) Q_{ac}^2 ;
\label{eq1c}
\end{align}

\vspace{0.7cm}

\begin{align}
& I_{100}^{\sigma\pi_s} \propto (Q_{ac}^2 + m_b^2) \cos^2\theta_B \cos^2\psi ; \nonumber \\
& I_{100}^{\sigma\sigma} = 0 ;
\label{eq2c}
\end{align}

\vspace{0.7cm}

\begin{align}
& I_{103}^{\sigma\pi_s} & \propto (0.535 \cos(2\psi) +0.44 \sin\psi)^2 Q_{ab}^2 \nonumber \\
&														  & + (0.535 -0.44 \sin\psi)^2m_c^2 ; \nonumber \\
&I_{103}^{\sigma\sigma} & \propto 0.56 \sin^2(2\psi) Q_{ab}^2 ; 
\label{eq3c}
\end{align}

\vspace{0.7cm}

\begin{align}
& I_{010}^{\sigma\pi_s} \propto  (Q_{ab}^2 + m_c^2) \cos^2\theta_B \cos^2\psi ; \nonumber \\
& I_{010}^{\sigma\sigma} =0 ;
\label{eq4c}
\end{align}

\vspace{0.7cm}

\begin{align}
& I_{110}^{\sigma\pi_s} & \propto  (0.37 \cos(2\psi) +0.61 \cos\psi)^2 Q_{bc}^2 \nonumber \\
&													&	 +  (0.37 +0.61 \cos\psi)^2 m_a^2 ; \nonumber \\
& I_{110}^{\sigma\sigma} & \propto 0.48 \sin^2(2\psi) Q_{bc}^2 ;
\label{eq5c}
\end{align}

\vspace{0.7cm}

\begin{align}
& I_{003}^{\sigma\pi_s} \propto (Q_{bc}^2 + m_a^2) \cos^2\theta_B \sin^2\psi \nonumber \\
& I_{003}^{\sigma\sigma} = 0.
\label{eq6c}
\end{align}

As stated in the main text, the $\sigma\sigma$ channel is zero for all on-axis reflections. It is therefore not possible to disentangle magnetic and orbital OP at these reflections where they have the same azimuth dependence in the $\sigma\pi$ channel. For the same reasons, it is not possible to infer, from the equality of the $\sigma\pi$ scattering with the total scattering, that these reflections are purely magnetic.

\section{REXS experimental results and numerical FDMNES analysis}

\begin{figure}[ht]
	\centering
	\includegraphics[width=0.4\textwidth]{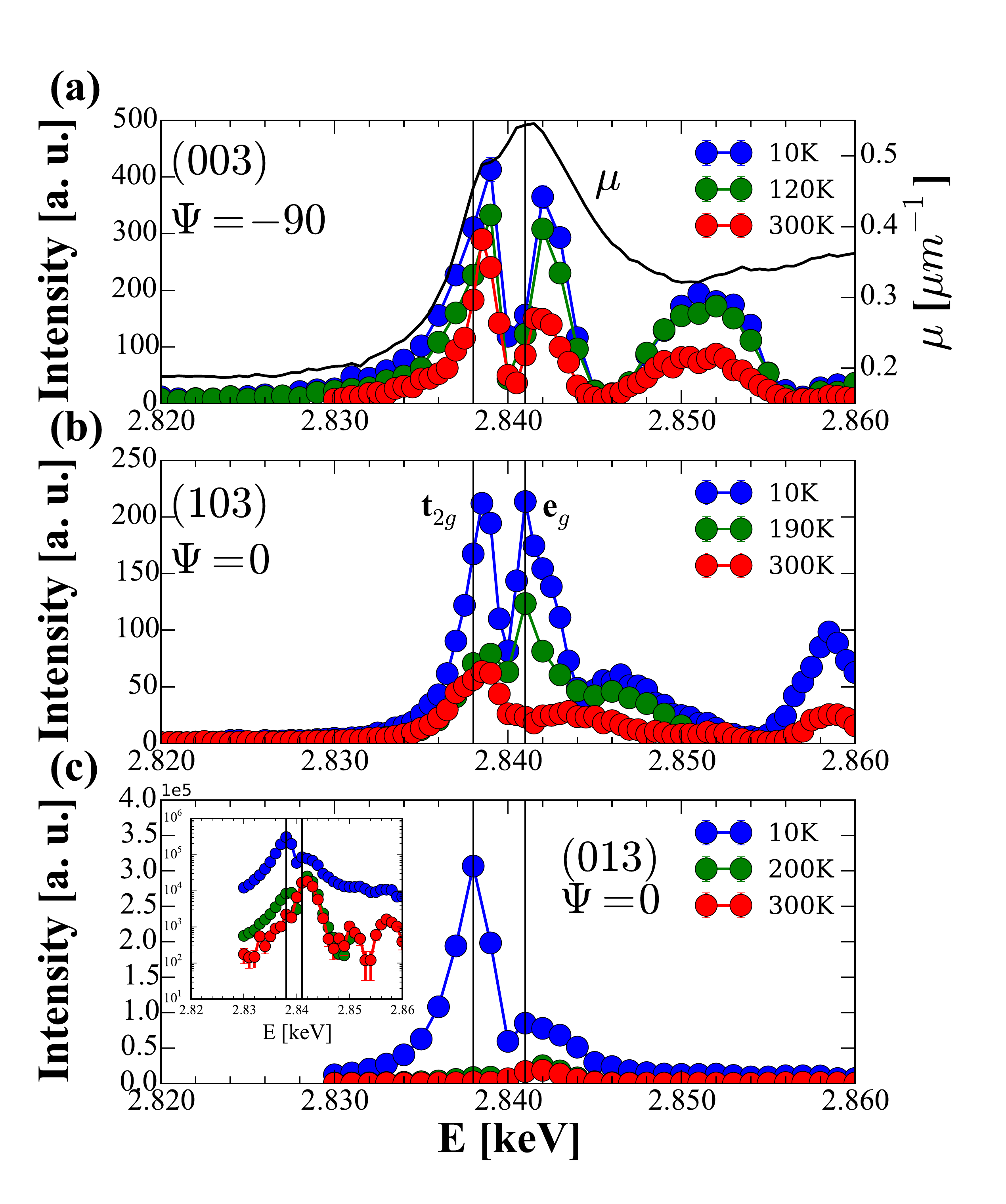} 
	\caption{(Colour online) Dependence of resonant intensity at several reflections at the Ru L$_3$ absorption edge. Each measurement is made at the specified azimuth using an area detector, summing both $\sigma\sigma$ and $\sigma\pi$ channels and corrected for self-absorption. The absorption cross-section, $\mu$, is also shown in (a), determined from the fluorescence. From it, the approximate position of t$_{2g}$ and e$_g$ orbitals is highlighted by vertical lines.  The log scale inset in (c) indicates the remaining resonant intensity above the magnetic transition on the (013) reflection.} \label{fig:xrayl3}
\end{figure}

\begin{figure}[ht]
	\centering
	\includegraphics[width=0.4\textwidth]{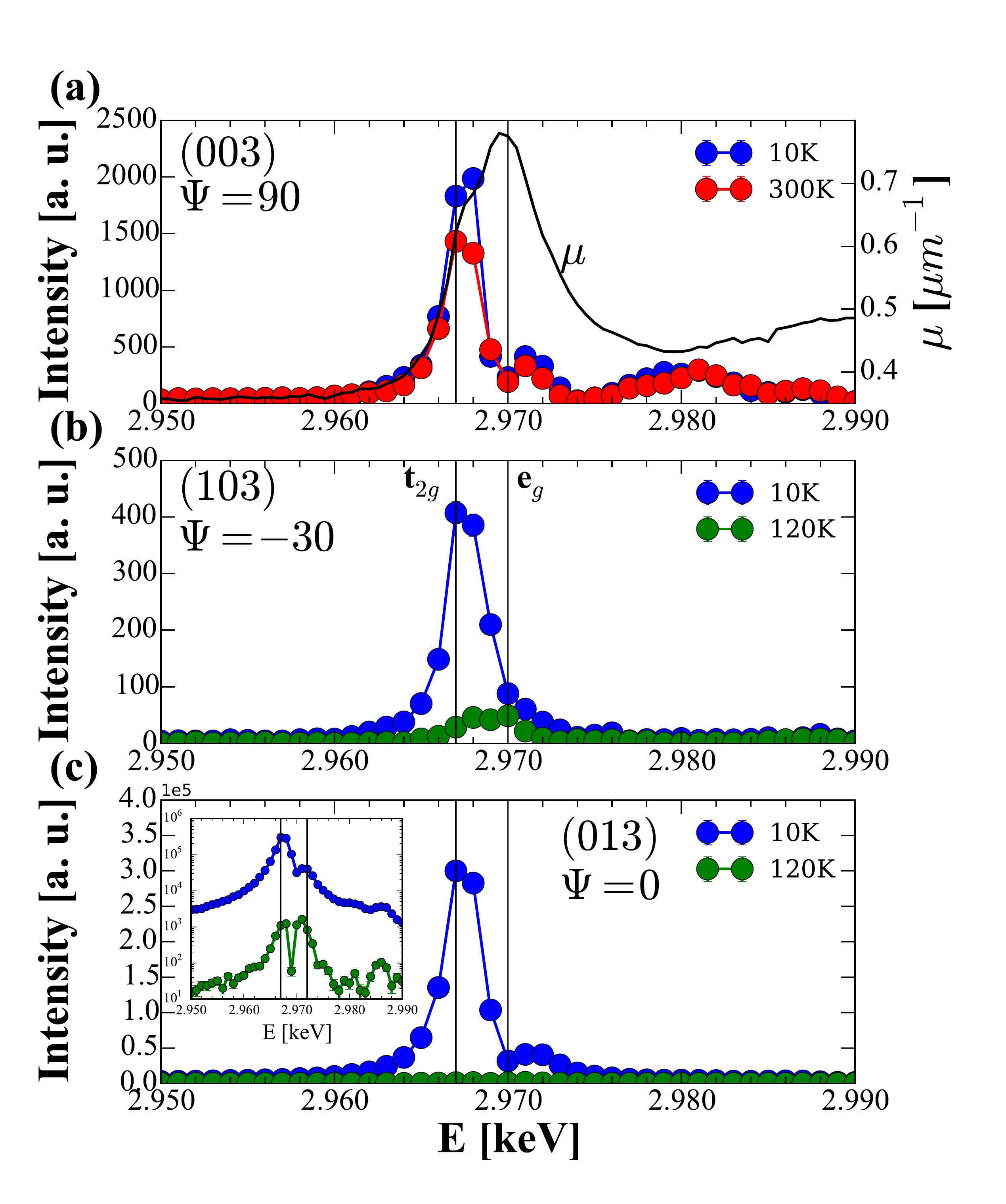} 
	\caption{(Colour online) Dependence of resonant intensity at several reflections at the Ru L$_2$ absorption edge. Each measurement is made at the specified azimuth using an area detector, summing both $\sigma\sigma$ and $\sigma\pi$ channels and corrected for self-absorption. The absorption cross-section, $\mu$, is also shown in (a), determined from the fluorescence. From it, the approximate position of t$_{2g}$ and e$_g$ orbitals is highlighted by vertical lines.  The log scale inset in (c) indicates the remaining resonant intensity above the magnetic transition on the (013) reflection.} \label{fig:xrayl2}
\end{figure}

Resonant x-ray diffraction is particularly sensitive to electronic and magnetic ordering through the excitation of core-state electrons into unoccupied electronic states. In the case of the Ru L$_{2,3}$ edges these unoccupied states are 4d orbitals. Hence resonant scattering is potentially sensitive to the preferential occupancy of specific 4d orbitals. Resonant reflections were only found for a propagation vector $\tau=\left(0,0,0\right)$ in the \cro{} lattice, indicating no period doubling or incommensurate ordering. Figures \ref{fig:xrayl3} and \ref{fig:xrayl2} show the resonant spectra observed in the rotated $\sigma\pi$ channel at three $\vec{Q}$-values matching the extinction rules of the Pbca space group at the Ru L$_3$ (2.828 keV) and L$_2$ edges (2.967 keV), respectively. At all reflections, two principal resonances are observed around the resonant edge, where the lower peak associates with \ttg{} excitations and the higher energy peak associates with \eg{} excitations. A few of the reflections show an additional broad resonance at higher energy, around 10 eV above the edge, whose origin could not be attributed by our numerical simulations. 

Each of the reflections shown is sensitive to a different projection of the magnetic and orbital components, indicated in Table \ref{tab:ref_sym}. The temperature dependence of these different energy spectra is shown as overlapping lines in Figs. \ref{fig:xrayl3} and \ref{fig:xrayl2}. 

\begin{figure}[ht]
	\centering
	\includegraphics[width=0.4\textwidth]{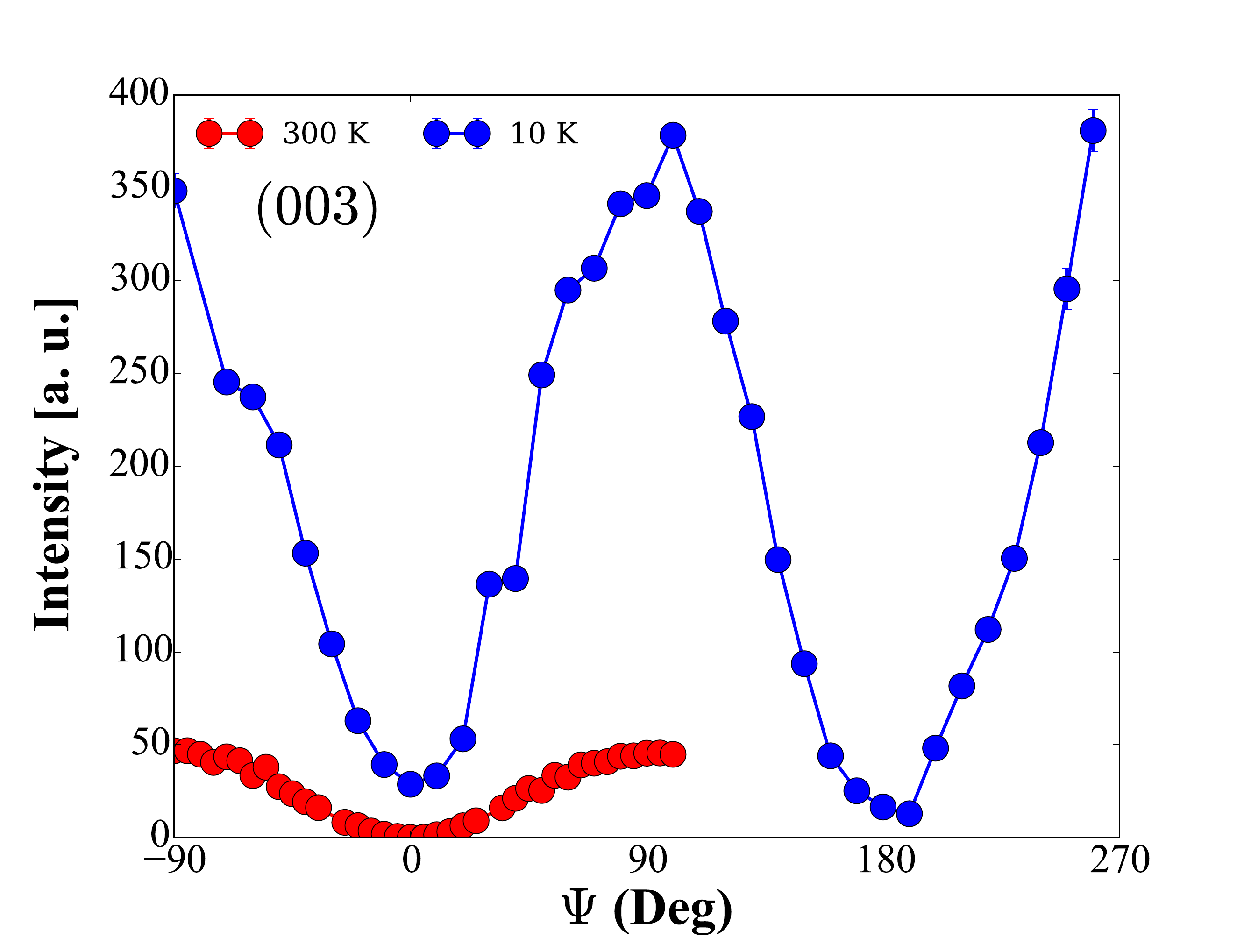} 
	\caption{(Colour online) Variation in intensity of (003) resonant reflection with azimuthal angle in $\sigma\pi$ rotated channel. The azimuthal reference is along the $(010)$, meaning that the angle is zero when the b-axis is within the scattering plane of the diffractometer.} \label{fig:xrayazi}
\end{figure}

To determine the azimuthal dependence of the reflections, two separate samples were aligned with either the ${001}$ or ${100}$ directions within the scattering plane of the diffractometer, allowing rotations about the azimuth of reflections $(003)$ and $(100)$ respectively. The intensity variation in each case is presented in Fig. \ref{fig:xrayazi} for the (003) and in Fig. \ref{fig:xrayazi100} for the (100). In both cases the azimuthal variations exhibit a 180-degree periodicity although the maxima are out of phase by 90-degrees between the two reflections, in agreement with Eqs. \ref{eq6c} and \ref{eq2c}, respectively. This periodicity is independent of temperature in both reflections, only the maximum height depends on temperature.

\begin{figure}[ht]
	\centering
	\includegraphics[width=0.4\textwidth]{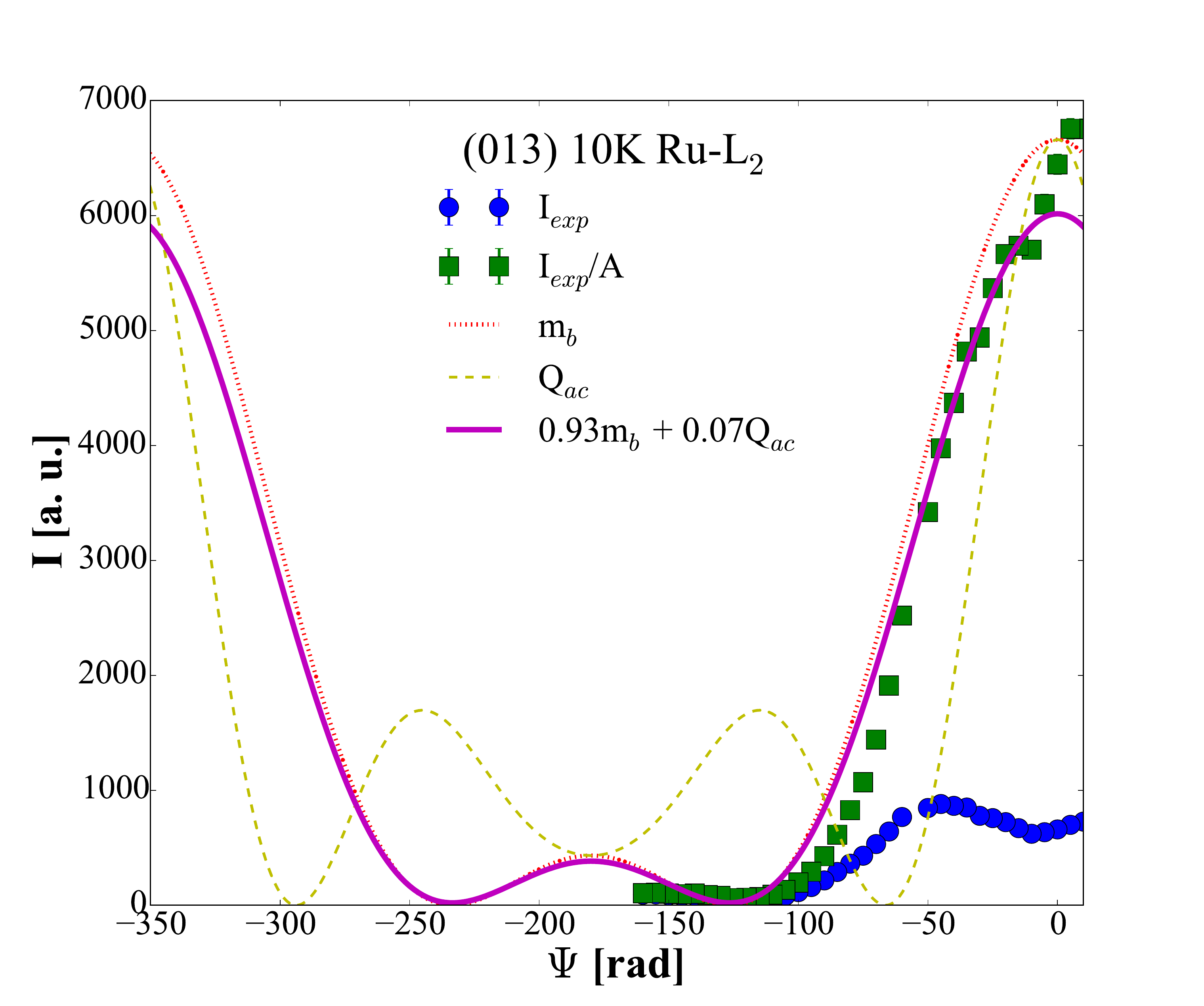} 
	\caption{(Colour online) Variation in intensity around the azimuth of the $(013)$ off-specular reflection. Both the as-measured and absorption corrected intensities are shown, highlighting the significance of the correction. The corrected intensity is very close to the signal expected from a pure magnetic moment along the $m_b$ direction.} \label{fig:xrayazi2}
\end{figure}

As described in Appendix C above, the azimuthal periodicity of any reflection along a principal direction is not dependent on the source of the scattering (whether magnetic or orbital) and cannot be used to separate magnetic and orbital contributions to the scattering. Reflections that are not along principal directions are instead separately sensitive to the magnetic moments and or orbital anisotropies and produce distinctly different forms depending on the origin of the signal. Measurements of the off-specular reflection $(013)$ were taken from a sample with a ${001}$ surface. To correct for self-absorption, the measured intensities are divided by the factor:

\begin{align}
A(\mathbf{Q},\psi) = \frac{1}{\mu}\left[1+\frac{\sin(\phi_1(\mathbf{Q},\psi))}{\sin(\phi_2(\mathbf{Q},\psi))}\right]^{-1}
\end{align} 

\noindent where $\mu$ is the absorption coefficient and the the angles $\phi_1$ and $\phi_2$ are the entrance and exiting angles to the sample surface, respectively. The measured and corrected azimuthal dependence of the $(013)$ reflection is given in Fig. \ref{fig:xrayazi2}, where we also show the expected shape for magnetic ($m_b$) and orbital ($Q_{bc}$) scattering. The form of the measured azimuthal scan can be very well fitted by an almost purely magnetic signal, in keeping with the conclusions of Subsection IV.A.

\begin{figure}[ht]
	\centering
	\includegraphics[width=0.4\textwidth]{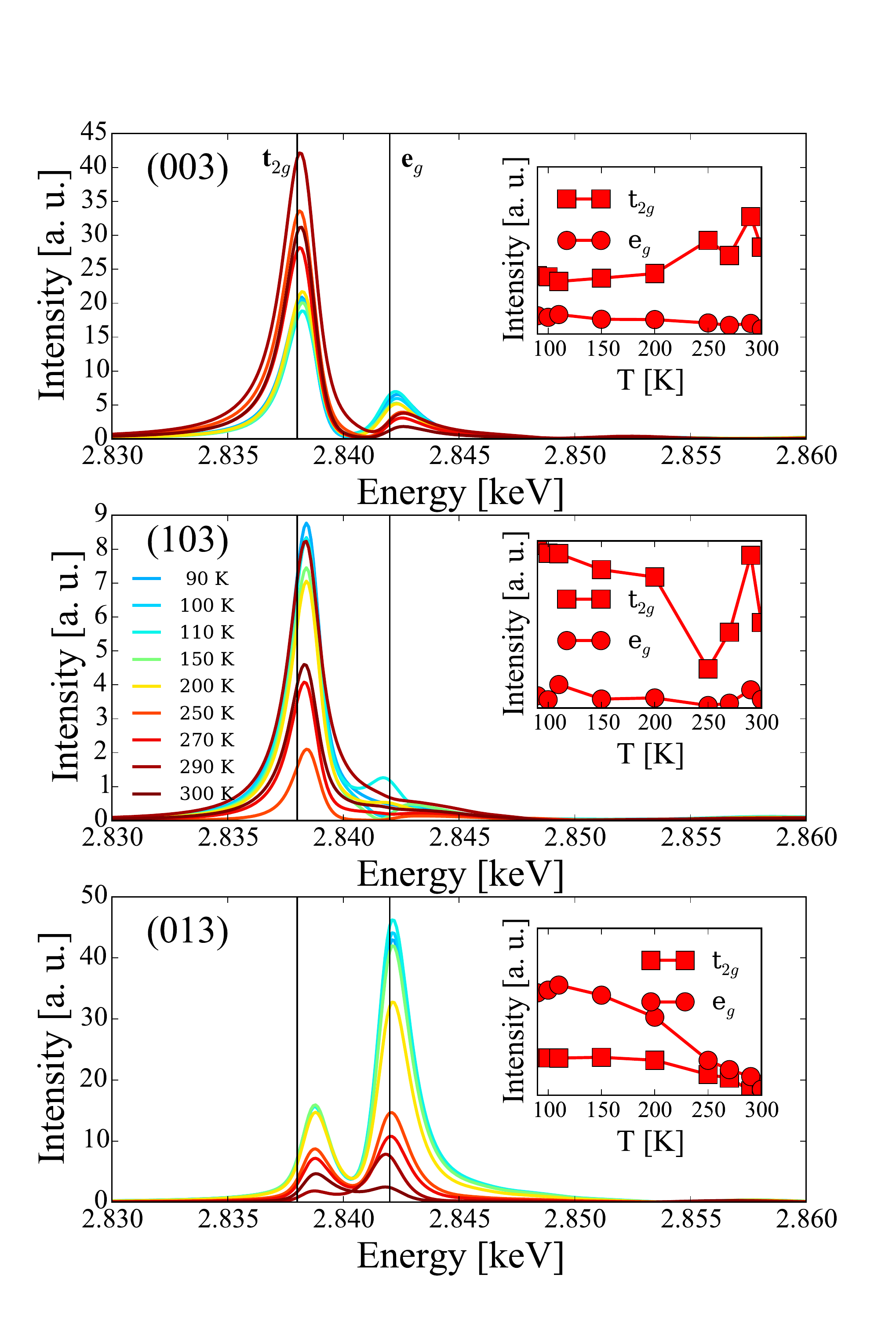} 
	\caption{(Colour online) Calculations of the resonant scattering arising solely from the effect of the atomic structure, as calculated by the FDMNES code. The $\sigma\pi$ resonant spectra of the three measured reflections is shown as light-to-dark lines for low-to-high temperatures, respectively, with the calculated intensity at the t$_{2g}$ and e$_g$ energies given inset.} \label{fig:fdmnes}
\end{figure}

Numerical calculations of the resonant spectra were made using the \textit{ab initio} FDMNES code\cite{Bunau2009}. The calculations were performed in the dipolar (E1-E1) approximation using the atomic coordinates refined from single crystal x-ray diffraction in Section 3. No magnetic or orbital ordering was included in the calculation. In this way we can infer that the resonant spectra produced are purely a result of the anisotropic tensor of susceptibility, or ATS scattering as it is commonly known. The calculated resonant scattering spectra for the main reflections measured are shown in Fig. \ref{fig:fdmnes}. Above the magnetic and orbital transitions, there is good qualitative agreement between the calculations and the measured spectra in Fig. \ref{fig:xrayl3}. The temperature dependence of these spectra, only due to the changing atomic coordinates of oxygen and calcium, shows a gradual increase in peak height with decreasing temperature. Such an increase is qualitatively similar to the increase seen at the (013) or (100) reflections, and yet not observed at the (103) or (003) reflections. In particular, no sharp change in peak height similar to the intensity change for the (100) reflection at 260K was reproduced. Again, as for the main conclusion of Subsection IV.B, this points to a physical mechanism that is not purely an effect of atomic rearrangements and distortions.

\bibliographystyle{apsrev4-1}
\bibliography{Ca2RuO4_biblio}

\end{document}